\documentclass[twocolumn]{aastex62}

\usepackage{tabularx}
\usepackage{rotating}
\usepackage{amsmath}

\graphicspath{{./}{figures/}}

\received{2019 July 26}
\revised{2020 January 20}
\accepted{2020 February 8}
\shorttitle{CO in 5 SF galaxies across the MS at z$\sim3.2$}
\shortauthors{Cassata et al.}
\begin{document}

\title{ALMA reveals the molecular gas properties of 5 star-forming 
   galaxies across the main sequence at $3<z<3.5$.}

\correspondingauthor{Paolo Cassata}
\email{paolo.cassata@unipd.it}

\author{Paolo Cassata}
\affiliation{Dipartimento di Fisica e Astronomia, Universit\`a di Padova, Vicolo dell'Osservatorio, 3 35122 Padova, Italy}
\affiliation{INAF Osservatorio Astronomico di Padova, vicolo dell'Osservatorio 5, I-35122 Padova, Italy}

\author{Daizhong Liu}
\affiliation{Max-Planck-Institut f\"{u}r Astronomie, Königstuhl 17, 69117 Heidelberg, Germany}

\author{Brent Groves}
\affiliation{Research School of Astronomy and Astrophysics, The Australian National University, Canberra, ACT 2611, Australia}

\author{Eva Schinnerer}
\affiliation{Max-Planck-Institut f\"{u}r Astronomie, Königstuhl 17, 69117 Heidelberg, Germany}

\author{Eduardo Ibar}
\affiliation{Instituto de F\'isica y Astronom\'ia, Universidad de Valpara\'iso, Avda. Gran Breta\~{n}a 1111, Valpara\'iso, Chile}

\author{Mark Sargent}
\affiliation{Astronomy Centre, Department of Physics and Astronomy, University of Sussex, Brighton BN1 9QH, UK}

\author{Alexander Karim}
\affiliation{Argelander-Institut f\"{u}r Astronomie, Auf dem Hügel 71, D-53121 Bonn, Germany}

\author{Margherita Talia}
\affiliation{Dipartimento di Fisica e Astronomia, Universit\'a di Bologna, Via Gobetti 93/2, I-40129 Bologna, Italy}
\affiliation{INAF-Osservatorio Astronomico di Bologna, Via Gobetti 93/3, I-40129, Bologna, Italy}

\author{Olivier Le F\`evre}
\affiliation{Aix Marseille Universit\'e, CNRS, LAM (Laboratoire d’Astrophysique de Marseille) UMR 7326, 13388, Marseille, France}

\author{Lidia Tasca}
\affiliation{Aix Marseille Universit\'e, CNRS, LAM (Laboratoire d’Astrophysique de Marseille) UMR 7326, 13388, Marseille, France}

\author{Brian C. Lemaux}
\affiliation{Department of Physics, University of California, Davis, One Shields Ave., Davis, CA 95616, USA}

\author{Bruno Ribeiro}
\affiliation{Leiden Observatory, Leiden University, PO Box 9513, 2300 RA, Leiden, The Netherlands}

\author{Stefano Fiore}
\affiliation{Dipartimento di Fisica e Astronomia, Universit\`a di Padova, Vicolo dell'Osservatorio, 3 35122 Padova, Italy}

\author{Michael Romano}
\affiliation{Dipartimento di Fisica e Astronomia, Universit\`a di Padova, Vicolo dell'Osservatorio, 3 35122 Padova, Italy}

\author{Chiara Mancini}
\affiliation{Dipartimento di Fisica e Astronomia, Universit\`a di Padova, Vicolo dell'Osservatorio, 3 35122 Padova, Italy}

\author{Laura Morselli}
\affiliation{Dipartimento di Fisica e Astronomia, Universit\`a di Padova, Vicolo dell'Osservatorio, 3 35122 Padova, Italy}

\author{Giulia Rodighiero}
\affiliation{Dipartimento di Fisica e Astronomia, Universit\`a di Padova, Vicolo dell'Osservatorio, 3 35122 Padova, Italy}
\affiliation{INAF Osservatorio Astronomico di Padova, vicolo dell'Osservatorio 5, I-35122 Padova, Italy}

\author{Luc\'ia Rodr\'iguez-Mu\~{n}oz}
\affiliation{Dipartimento di Fisica e Astronomia, Universit\`a di Padova, Vicolo dell'Osservatorio, 3 35122 Padova, Italy}

\author{Andrea Enia}
\affiliation{Dipartimento di Fisica e Astronomia, Universit\`a di Padova, Vicolo dell'Osservatorio, 3 35122 Padova, Italy}

\author{Vernesa Smolcic}
\affiliation{Department of Physics, Faculty of Science, University of Zagreb, Bijeni\v{c}ka cesta 32, 10000 Zagreb, Croatia}

%% Note that the \and command from previous versions of AASTeX is now
%% depreciated in this version as it is no longer necessary. AASTeX 
%% automatically takes care of all commas and "and"s between authors names.

%% AASTeX 6.2 has the new \collaboration and \nocollaboration commands to
%% provide the collaboration status of a group of authors. These commands 
%% can be used either before or after the list of corresponding authors. The
%% argument for \collaboration is the collaboration identifier. Authors are
%% encouraged to surround collaboration identifiers with ()s. The 
%% \nocollaboration command takes no argument and exists to indicate that
%% the nearby authors are not part of surrounding collaborations.

%% Mark off the abstract in the ``abstract'' environment. 
\begin{abstract}

We present the detection of CO(5-4) with S/N$>7-13$ and a lower CO transition with S/N$>3$ (CO(4-3) for 4 galaxies, and CO(3-2) for one) with ALMA in band 3 and 4 in five main sequence star-forming galaxies with stellar masses $3-6\times 10^{10} M/M_{\odot}$ at $3<z<3.5$. We find a good correlation between the total far-infrared luminosity $L_{FIR}$ and the luminosity of the CO(5-4) transition $L'_{CO(5-4)}$, where $L'_{CO(5-4)}$ increases with SFR, indicating that CO(5-4) is a good tracer of the obscured SFR in these galaxies. The two galaxies that lie closer to the star-forming main sequence have CO SLED slopes that are comparable to other star-forming populations, such as local SMGs and BzK star-forming galaxies; the three objects with higher specific star formation rates (sSFR) have far steeper CO SLEDs, which possibly indicates a more concentrated episode of star formation. By exploiting the CO SLED slopes to extrapolate the luminosity of the CO(1-0) transition, and using a classical conversion factor for main sequence galaxies of $\alpha_{CO}=3.8~M_{\odot} (K~km~s^{-1}~pc^{-2})^{-1} $, we find that these galaxies are very gas rich, with molecular gas fractions between 60 and 80\%, and quite long depletion times, between 0.2 and 1 Gyr. Finally, we obtain dynamical masses that are comparable with the sum of stellar and gas mass (at least for four out of five galaxies), allowing us to put a first constraint on the $\alpha_{CO}$ parameter for main sequence galaxies at an unprecedented redshift.

\end{abstract}

%% Keywords should appear after the \end{abstract} command. 
%% See the online documentation for the full list of available subject
%% keywords and the rules for their use.

%% From the front matter, we move on to the body of the paper.
%% Sections are demarcated by \section and \subsection, respectively.
%% Observe the use of the LaTeX \label
%% command after the \subsection to give a symbolic KEY to the
%% subsection for cross-referencing in a \ref command.
%% You can use LaTeX's \ref and \label commands to keep track of
%% cross-references to sections, equations, tables, and figures.
%% That way, if you change the order of any elements, LaTeX will
%% automatically renumber them.
%%
%% We recommend that authors also use the natbib \citep
%% and \citet commands to identify citations.  The citations are
%% tied to the reference list via symbolic KEYs. The KEY corresponds
%% to the KEY in the \bibitem in the reference list below. 

\section{Introduction}

The global star formation rate density of the Universe (SFRD) increased by a factor of $\sim$15 in the first 3 Gyr after the Big Bang, and then fell off by a similar factor down to the local Universe (Cucciati et al. 2012, Madau~\&~Dickinson 2014). There is a wide consensus nowadays that the bulk of the star formation activity at all epochs occurs in galaxies that lie on a relatively narrow sequence in the stellar mass vs SFR plane (main sequence, MS; Noeske et al. 2007; Daddi et al. 2007; Rodighiero et al. 2011). Galaxies were forming stars faster at earlier epochs: at z$\sim$3 the sSFR (SFR/$M_{\odot}$) of galaxies on the MS is $\sim$100 times higher than local galaxies (Faisst et al. 2016, Tasca et al. 2015; Schreiber et al. 2016; Tomczak et al. 2016; Leslie et al. 2019). Despite knowing when and where (i.e. mainly in $\log(M_*)\sim3-10\times10^{10} M_{\odot}$ MS galaxies) the stars in the Universe have formed, the mechanisms that trigger such rapid increase in the global star formation activity of the Universe are still debated: is it due to an increase in the gas fraction or a higher star-formation efficiency, or a combination of the two (e.g. Geach et al. 2011, Saintonge et al. 2013, Genzel et al. 2015, Scoville et al. 2016)? In order to make significant progress we need to link the nature of the star-forming galaxies on the MS with their gas reservoirs, eventually probing the efficiency of the star formation and the gas fraction and their evolution with cosmic time.

Although the molecular gas mass can be estimated from rather inexpensive dust continuum detections (Hildebrand et al. 1983; Magdis et al. 2012a; Scoville et al. 2014, Scoville et al. 2016; Groves et al. 2015), a more direct tracer of the properties of the molecular gas and of the total molecular mass is the CO line (Carilli~\&~Walter 2013). CO detections are indeed quite standard for MS galaxies at $z < 2$ (Saintonge et al. 2011; Genzel et al. 2015; Villanueva et al. 2017; Tacconi et al. 2018), but they require increasingly longer integration times at $2 < z < 6$, the critical epoch when the SFRD of the Universe experienced its accelerated growth. Therefore, in comparison to the few hundreds of MS star-forming galaxies with CO detection at $z < 1$, and the few tens that are detected at $1 < z < 3$ (Daddi et al. 2008; Daddi et al. 2010; Genzel et al. 2015; Tacconi et al. 2018; Sharon et al. 2016), only a handful of MS galaxies at $z > 2$ have been detected in CO: Magdis et al. (2012b) obtained CO(3-2) fluxes for 2 Lyman-break galaxies at $z \sim 3$, and Saintonge et al. (2013) and Dessauges-Zavadsky et al. (2017) detected CO for 4 lensed main sequence galaxies at $2.7 < z < 3$ and one at $z\sim$ 3.6 (but with additional uncertainties due to the uncertain magnification factor), respectively.

As a result, our knowledge about the ISM mass and properties at these epochs are based on sub-mm continuum detections (e.g. Schinnerer et al. 2016, Liu et al. 2019) that are subject to assumptions that are difficult to control (e.g. on the dust temperature and on the evolving metallicity at $z > 3$). In order to make progress, we observed during ALMA Cycle 3 and 4 a sample of five main sequence galaxies at $3<z<3.5$ targeting two CO transitions (CO(5-4) for the whole sample, CO(4-3) for 4 galaxies and CO(3-2) for one), in order to constrain their ISM masses using the CO emission. This technique in fact relies on different assumptions (e.g. on the CO excitation state and gas density, see Narayanan~\&~Krumholz~2014, but also again on metallicity), and therefore provides an independent estimate of ISM mass. Throughout this work, we assume $H_0=70 km~s^{-1} Mpc^{-1}$, $\Omega_{\Lambda} = 0.7, \Omega_M = 0.3$. We use stellar masses and SFRs based on a Chabrier IMF (or converted to a Chabrier IMF, when necessary).

%__________________________________________________________________

\section{Sample and data}

We selected the galaxies for this work starting from the sample in Schinnerer et al. (2016), by choosing the five sources for which we estimated the brightest CO(5-4) flux, based on their dust detections at 240 GHz in ALMA band 6, among those for which a spectroscopic redshift was available as a part of the VIMOS Ultra-Deep Survey (VUDS; Le F\`evre et al. 2015) and of the zCOSMOS Deep survey (Lilly et al. 2007).

   \begin{figure}
   \centering
  \includegraphics[width=\columnwidth]{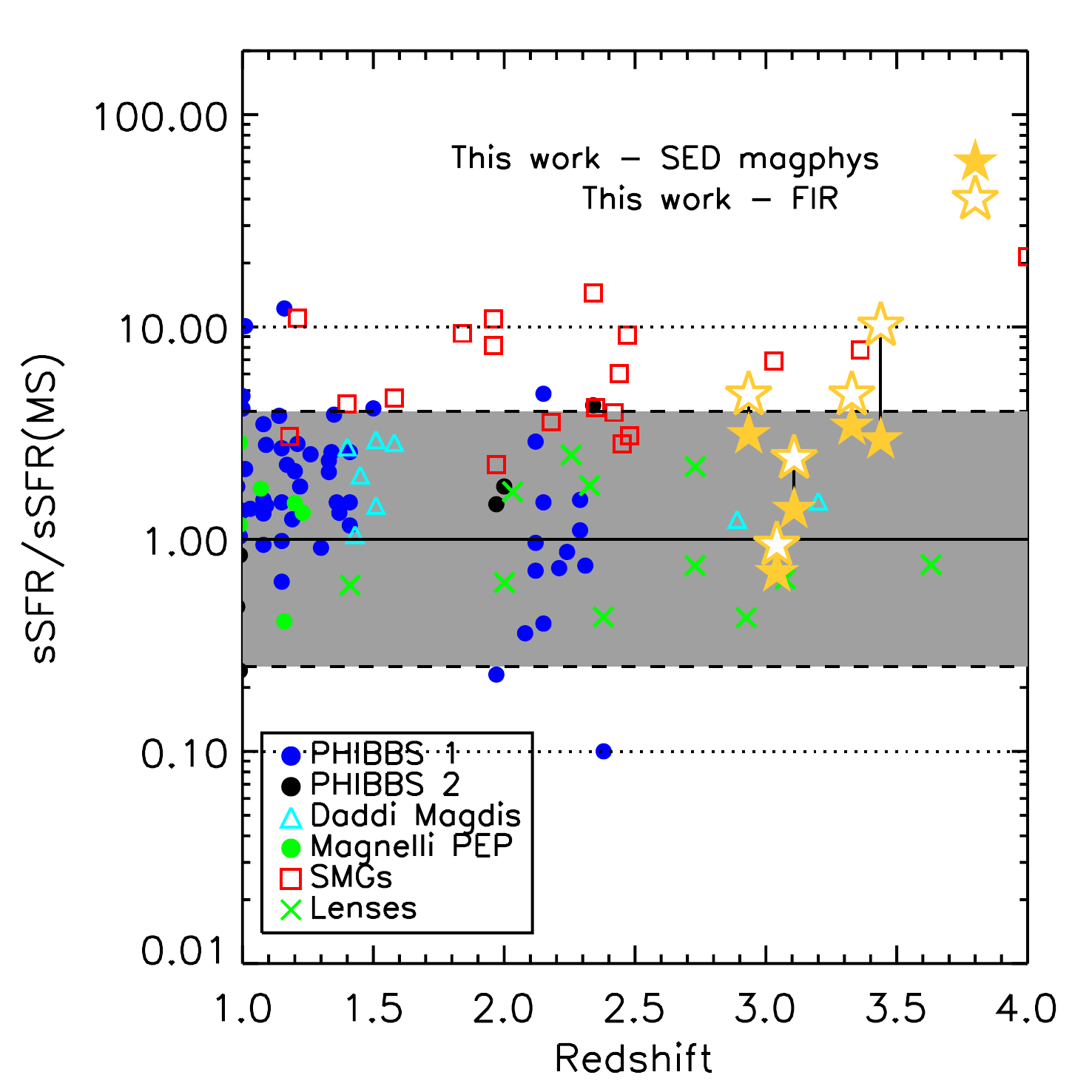}
     \caption{sSFR, normalized to the one estimated for the main sequence (sSFR(MS)) as a function of redshift, for different samples of normal "main sequence" galaxies with at least one CO line detected in their spectra. Blue points and black points show, respectively, galaxies from the PHIBBS 1 survey (Tacconi et al. 2010, 2013), and PHIBBS 2 survey (Combes et al. 2016); red squares indicate SMGs from Greve et al. (2005), Tacconi et al. (2006), Tacconi et al. (2008), Bothwell et al. (2013); cyan triangles indicate star-forming BzK objects from Daddi et al. (2010) and Magdis et al. (2012); filled green circles are galaxies from Magnelli et al. (2012); green crosses indicate lensed galaxies from Saintonge et al. (2013) and Dessauges-Zavadsky et al. (2017). Yellow stars show the five objects presented in this work: filled stars indicate the SFR that is obtained by the {\rm MAGPHYS} tool; empty stars indicate the star formation rates that are obtained converting the $L_{FIR}$ using the classical Kennicutt relation. The horizontal continuous line marks the location of the MS, the dashed (dotted) lines indicate  the locii 4x (10x) times above or below the MS. The grey area emphasizes the location of the main sequence, according to the classical definition by Rodighiero et al. (2011; 2014), and Elbaz et al. (2018). }
              \label{Fig:deltams}%
    \end{figure}

\subsection{Ancillary data and SED fitting}
\begin{table*}[!ht]
    \centering
\begin{tabular}{c c c c c c c c c}%{c|c|c|c|c|c}
\hline\hline \noalign{\smallskip} 
     & RA & DEC & $z_{spec}$ & $\log(M_*)$ & $\log(SFR)^a$ & $\log(L_{FIR})$ & $T_{dust}$ & $\delta_{MS}^b$\\
            &  & & & [$M_{\odot}$] & [$M_{\odot} yr^{-1}$] & $L_{\odot}$ & K\\
 \noalign{\smallskip} \hline
\noalign{\smallskip}

 Gal1 & 10:01:23.182 &   +02:36:26.06 & 3.1120$^{\dagger}$ & 10.86 & 2.49(2.73) & 12.73 & 54.1$\pm$6.3 & 1.34(2.40) \\
 Gal2 & 09:59:38.292 &   +02:13:19.93 & 3.0494$^{\dagger}$ & 10.82 & 2.17(2.30) & 12.30 & 30.7$\pm$2.4 & 0.69(0.94) \\
 Gal3 & 10:01:19.546 &   +02:09:44.53 & 2.9342$^{\dagger}$ & 10.49 & 2.53(2.71) & 12.71 & 39.8$\pm$1.0 & 3.11(4.79) \\
 Gal4 & 10:01:06.802 &   +02:15:31.74 & 3.4388$^{\dagger\dagger}$ & 10.82 & 2.80(3.33) & 13.33 & 62.1$\pm$1.3 & 2.94(10.14)\\
 Gal5 & 09:59:30.523 &   +02:17:01.95 & 3.3428$^{\dagger}$ & 10.80 & 2.86(3.00) & 12.82 & 54.2$\pm$2.7 & 3.41(4.80) \\
\noalign{\smallskip} \hline
\noalign{\smallskip}
 
     \end{tabular}
\tablenotetext{\dagger}{based on ISM lines}
\tablenotetext{\dagger\dagger}{~based on Lyman-$\alpha$}
\tablenotetext{a}{The first value is the SFR estimated by MAGPHYS, the one in parenthesis is based on $L_{FIR}$}
\tablenotetext{b}{Defined as $sSFR/sSFR(MS)$; first value based on MAGPHYS, the one in parenthesis on $L_{FIR}$}

\caption{General properties of the sample}
    \label{tab:properties_photo}
\end{table*} 

The galaxies in this sample lie in the COSMOS field, one of the most widely studied patches of the sky. Multi-wavelength photometry from the UV to FIR rest-frame, including GALEX, CFHT, Subaru, VISTA, Spitzer, Herschel and VLA is available for the whole field, as part of the COSMOS survey (Scoville et al. 2007; Sanders et al. 2007; Capak et al. 2007; McCracken et al. 2012; Laigle et al. 2016). Spectroscopy is crucial to have robust and precise spectroscopic redshifts that are needed to tune the ALMA setup around the expected frequency of the CO lines: two galaxies have spectra from the VIMOS Ultra-Deep Survey (VUDS; Le F\`evre et al. 2015) and three from the zCOSMOS Deep survey (Lilly et al. 2007). The published redshifts for these galaxies are based on cross-matching techniques that fit the positions of the Lyman$\alpha$ line and of the ISM lines together, although these are almost always offset with respect to the systemic velocity by several hundreds kilometers in opposite directions (Steidel et al. 2010; Marchi et al. 2019; Cassata et al. in prep). Nevertheless, with the frequency coverage of the ALMA correlator (4 GHz in each of two sidebands) the CO line is expected to fall within the covered bands using either Ly$\alpha$ or the ISM lines, if those are not offset with respect to the systemic velocity by more than $\sim\pm2000$ km/s. However, we also manually re-measured the spectroscopic redshifts of the five galaxies, in order to compare them with the ones derived from the CO emission. When possible, we used the detected ISM lines to fix the redshift; in one case the ISM lines were too noisy and we used the bright Ly$\alpha$ emission instead. 

We fitted the rich multi-wavelength photometry in order to obtain physical information such as star formation rates, stellar masses, dust temperatures. For this work, we used the new "super-deblended" catalog by Jin et al. (2018), in which the emission in the FIR bands for each object is accurately deblended over multiple objects by using the position of the emission at shorter wavelength as a prior, in combination with UV-optical-NIR photometry from Laigle et al. (2016). By checking the multi-wavelength images, we realized that the photometry for one object, Gal2, is perturbed by a lower redshift interloper that is not deblended in the Laigle et al. (2016) catalog: in fact, this interloper is detected in the U-band image, a band that at z$\sim$3 would match the Lyman-continuum. We therefore performed a band by band manual deblending, defining two apertures, one around the lower redshift interloper, and the other around Gal2: we estimated, band by band, the fraction of the total flux for each of the two components, and we use the deblended values to build a Spectral Energy Distribution (SED) for Gal2.

All five galaxies are detected all the way from the UV rest-frame to the 24 $\mu$m band; four out of five are also detected with Herschel in the FIR rest-frame (around the peak of the cold dust thermal emission at $\sim$ 100 $\mu$m rest-frame). By construction, the continuum in ALMA band 6 at 240 GHz  is detected for all galaxies, and we included it in the SED fitting along with the continuum in band 4 band, that we obtained as a part of this project (see next Section for details). We fitted the multi-band photometry with {\it MAGPHYS} (da Cunha, Charlot ~\&~ Elbaz 2008), with the redshift fixed to the spectroscopic one, and including the photometric point from the ALMA band 4~\&~6 continuum observations. {\it MAGPHYS} fits the whole multi-wavelength spectral energy distribution from the UV to the FIR/sub-mm, ensuring the balance between the energy absorbed by the dust in the UV and re-emitted in the FIR. As outputs of the fitting procedure, {\it MAGPHYS} provides stellar masses, star formation rates, and total FIR luminosities $L_{FIR}$, among others. The galaxies turn out to be quite massive, with stellar masses in the range $3\times10^{10}<M_*/M_{\odot}<8\times10^{10}$, and star-forming, with SFR$\sim100-600 M_{\odot}~yr^{-1}$. As a check, we then derived the obscured SFR from the total FIR luminosities using the classical Kennicutt (1998) conversion, assuming a Chabrier IMF: this should be a good proxy of the total star formation rate, since all five objects are quite obscured at UV rest-frame wavelengths. However, we obtain from the FIR SFR$\sim200-2000 M_{\odot}~yr^{-1}$, in all cases larger, by a factor of 1.5-3.5, than the ones obtained from {\it MAGPHYS}. This is not completely unexpected, as old stellar populations can also heat up the dust in the ISM (da Cunha et al. 2010), and our galaxies are already quite massive at $z\sim3$, implying an underlying population of old stars, ontop of which new stars are forming. Therefore, we consider the SFRs from MAGPHYS more robust, and we prefer them in the remainder of the paper, but, to be conservative, we also show how the results would change should the Kennicutt law be used to compute the SFR. Coordinates, redshifts, stellar masses, SFRs, FIR luminosities, dust temperatures and distances from the main sequence $\delta_{MS}=sSFR/sSFR(MS)$ are summarized in Table~\ref{tab:properties_photo}.

Both from photometry and spectroscopy, the five objects have all the properties typical of normal star-forming galaxies: Ly$\alpha$ is bright in one galaxy while very weak or in absorption in the other four, ISM features such as OI, CII, CIV and SiIV are observed in absorption, and there is no evidence of AGN features in the observed optical spectra or of a warm torus in the broad band photometry.
\begin{table*}[!ht]
    \centering
\begin{tabular}{c c c c c c c}%{c|c|c|c|c|c}
\hline\hline \noalign{\smallskip} 
      & Beam band3 & Beam band4 & ToS$^a$ band 3 & ToS band 4 & rms$^b$ band 3 & rms$^b$ band 4\\
      & arcsec$\times$arcsec & arcsec$\times$arcsec  & min & min & mJy/beam & mJy/beam\\
  \noalign{\smallskip} \hline
\noalign{\smallskip}
Ga11 & 0.61$\times$0.85 & 0.76$\times$1.48 & 21.17 & 9.58 & 0.47 & 0.40\\
Ga12 & 0.61$\times$0.85 & 0.76$\times$1.48 & 21.17 & 9.58 & 0.73 & 0.50\\
Gal3 & 0.73$\times$0.92 & 0.81$\times$1.08 & 3.53  & 4.54 & 0.95 & 0.65\\
Gal4 & 0.61$\times$0.64 & 0.6$\times$0.72  & 11.59 & 25.20 & 0.58 & 0.34\\
Gal5 & 0.61$\times$0.64 & 0.6$\times$0.72  & 11.59 & 25.20 & 0.63 & 0.34\\
\noalign{\smallskip} \hline
\noalign{\smallskip}
 
     \end{tabular}
     \tablenotetext{a}{Time on Source}
     \tablenotetext{b}{the rms are calculated, in channels of 40 km/s, with the CASA tool IMSTAT in the cubes that are the result of the CLEAN process, described in the text.}
   \caption{Properties of the ALMA observations} 
    \label{tab:properties_alma_obs}
\end{table*} 

In Figure~\ref{Fig:deltams} we show their sSFR, relative to the specific star formation rate of galaxies on the main sequence sSFR(MS), defined using the MS presented by Schreiber et al. (2015) at the median redshift of $z = 3$) as a function of redshift, in comparison with similar samples of star-forming galaxies at $z > 1$ that have at least one CO transition detected in the sub-mm. The figure is built readapting Figure 1 from Villanueva et al. (2017) and Genzel et al. (2015), that used a slightly different parametrization of the MS, valid only up to $z\sim$ 2.5, by Whitaker et al. (2012): in any case, the two parametrizations do not differ too much in the stellar mass range spanned by the galaxies in this work, therefore the different samples can be compared consistently. From Fig. 1, the region $\pm$0.6dex around the MS is explored quite well up to redshift 2.5 by the galaxies from the PHIBBS 1~\&~2 samples (Tacconi et al. 2010, 2013; Combes et al. 2016), and by the samples presented in Magnelli et al. (2012), Daddi et al. (2010) and Magdis et al. (2012). Sub-mm galaxies (SMGs) samples from Greve et al. (2005), Tacconi et al. (2006), Tacconi et al. (2008), Bothwell et al. (2013) span a higher range of sSFR, typically 0.6dex above the MS. Only a handful of galaxies have CO detections at $z>3$: two galaxies from Daddi et al. (2010) and Magdis et al. (2012), 3 SMGs, and 2 lensed galaxies. For each galaxy in our sample we show in Fig 1 the two sSFR obtained from MAGPHYS and from rescaling the FIR luminosity (filled and empty yellow stars, respectively). It can be seen that our five galaxies span the upper half of the classical main sequence ($sSFR< 4\times sSFR(MS)$, Rodighiero et al. (2011; 2014); Elbaz et al. (2018)); however, two galaxies (Gal1~\&~Gal2) lie very close to the MS, while the other three sit close to the line that marks the transition between MS and starburst population (the SFR from MAGPHYS gives $sSFR < 4\times sSFR(MS)$, but the SFR from the FIR gives $sSFR > 4\times sSFR(MS)$).

\subsection{ALMA data}

  \begin{figure*}
   \centering
          \includegraphics[width=0.9\textwidth]{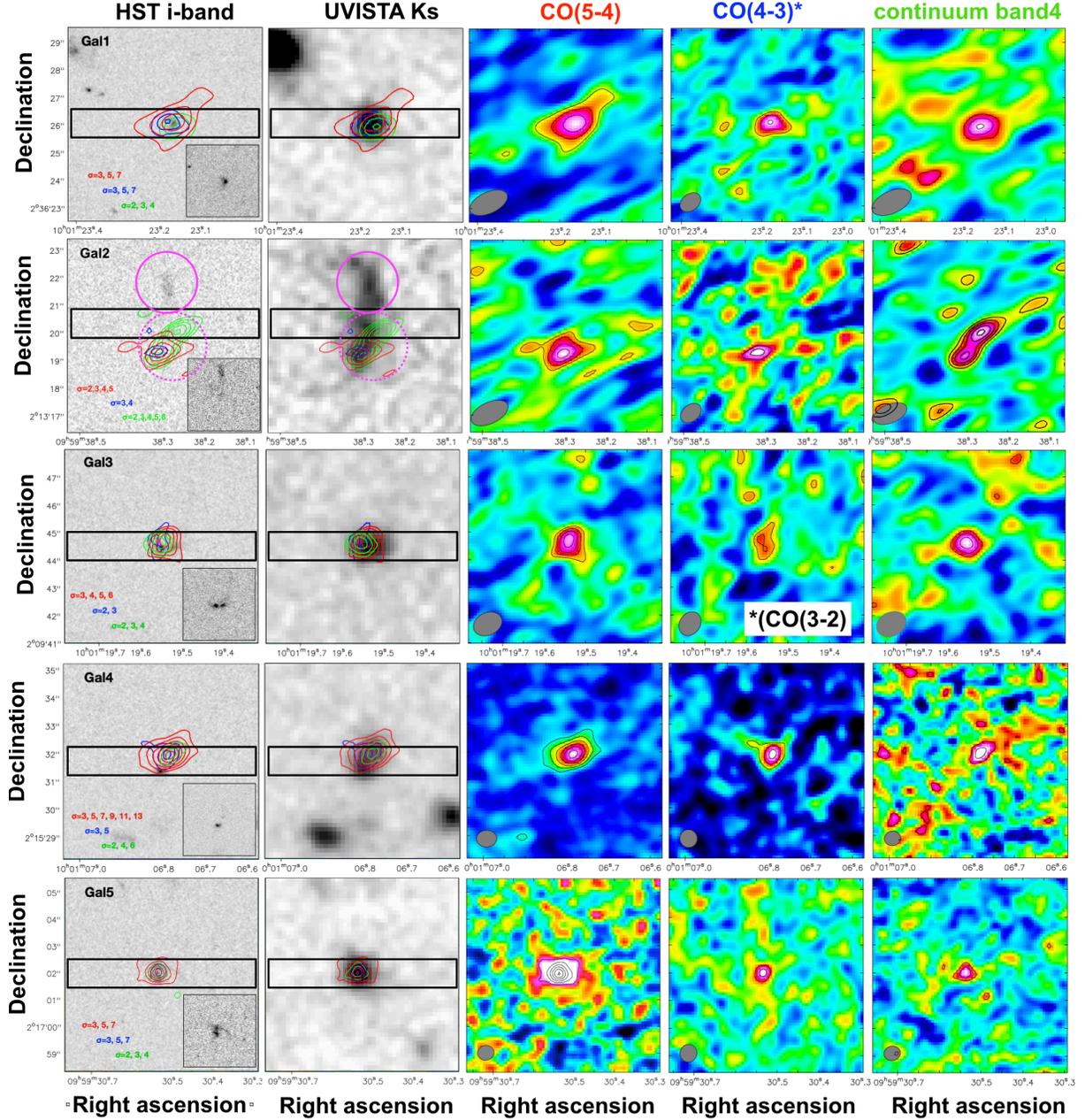}
     \caption{For each galaxy in the sample, we show, from left to right, the HST/F814W image from the COSMOS survey (Scoville et al. 2007, Koekemoer et al. 2007), the K$_s$ image from the UltraVISTA DR4 release (McCracken et al. 2012), with overlayed the contours of the CO(5-4) line (in red), the CO(4-3) or CO(3-2) (in blue) and the continuum in band 4 (in green); the CO(5-4) map in band 4; the CO(4-3) or CO(3-2) map in band 3; the continuum image in band 4 at $\sim500\mu$m. For the ALMA images, we used a natural weighting scheme, that maximizes the sensitivity of the maps; only for the band 4 continuum map of Gal2 we used a Briggs weigthing scheme, to resolve the extended emission. See text for details. The contours show increasing S/N steps, as indicated in each panel. The HST image is repeated, on the left, for each object, in a smaller panel without ALMA contours, in order to better show the UV rest-frame morphology. In the first two panels of each object the black rectangle shows the position of the spectroscopic slit.}
              \label{Fig:images}%
    \end{figure*}
    
The five galaxies in the sample were observed in ALMA Cycle 3 (2015.1.01590.S, PI: Cassata) in band 3 (around 110 GHz, configuration C40-5) and band 4 (around 140 GHz, configuration C40-4), between 9 of June 2016 and  31 of July 2016 (plus a repetition of a failed observation on 29 of October 2016) to target two CO transitions for each galaxy: CO(5-4) for all galaxies (in band 4), CO(4-3) for four galaxies and CO(3-2) for one (in band 3). The time spent on-source ranges from a few minutes to 25 minutes per object per band. Standard sources J1058+0133, J0948+0022 and Titan were used for calibration. 

The data analysis has been carried out with standard analysis pipelines available as part of CASA version 4.7 (McMullin et al. 2007). The cubes were cleaned and imaged adopting a natural weighting scheme, that maximizes the sensitivity to faint signal, and  using masks at source position and setting a threshold of 3$\times$rms noise level on the dirty images, that was measured to range from $\sim$0.35 to $\sim$0.95 mJy/beam. Although the natural resolution of the data is $\sim$ 20 km/s, we extracted the cubes in spectral bins of 40 km/s, which is more than enough to resolve lines that have spectral FWHM in excess of 200 km/s. The resulting clean beams have FWHM$\sim$0.6-1 arcesec (elliptical in a few cases). The clean beam sizes, on source times, and the noise levels of each image are reported in Table~\ref{tab:properties_alma_obs}: it can be seen that for Gal1~\&~Gal2 the beams in band 3 are quite smaller than those in band 4; for each the other 3 galaxies the beam size in band 3 matches quite well that in band 4. The largest recoverable angular scales are 6.7" and 7.5" for band 3 and band 4 observations, respectively, and therefore these configurations are appropriate to retrieve the total flux from objects that have diameters of the order of 2-3".

Initally, CO spectra are extracted from a circular region with 2" diameter around the positions of the expected emission. This first step is only used to identify the channels to integrate to obtain the moment 0 maps: we selected the channels above 1$\sigma$ of the cube rms bracketing the peak of the line emission. We then collapsed the cubes into moment 0 maps, with the {\it immoments} task. Continuum maps are instead obtained collapsing the 3 out of 4 sidebands not containing the CO emission. The maps are presented in Figure~\ref{Fig:images}, together with a HST image in the F814W filter from the COSMOS project (Koekemoer et al. 2007; Scoville et al. 2007). In order to obtain the total line and continuum fluxes, we first built a segmentation map, in which we keep all the pixels contiguous to the center in which the measured flux is above 2$\sigma$ (measured in a region not containing the source), after subtracting the continuum; we then integrate the flux from these pixels. This method includes less (noisy) pixels than a classical aperture photometry approach, and therefore maximizes the S/N of the measurements. Whenever the emission region is smaller than the clean beam, and therefore the emission is not resolved, we take the peak flux as the total flux.

\begin{figure}
   \centering
  \includegraphics[width=\columnwidth]{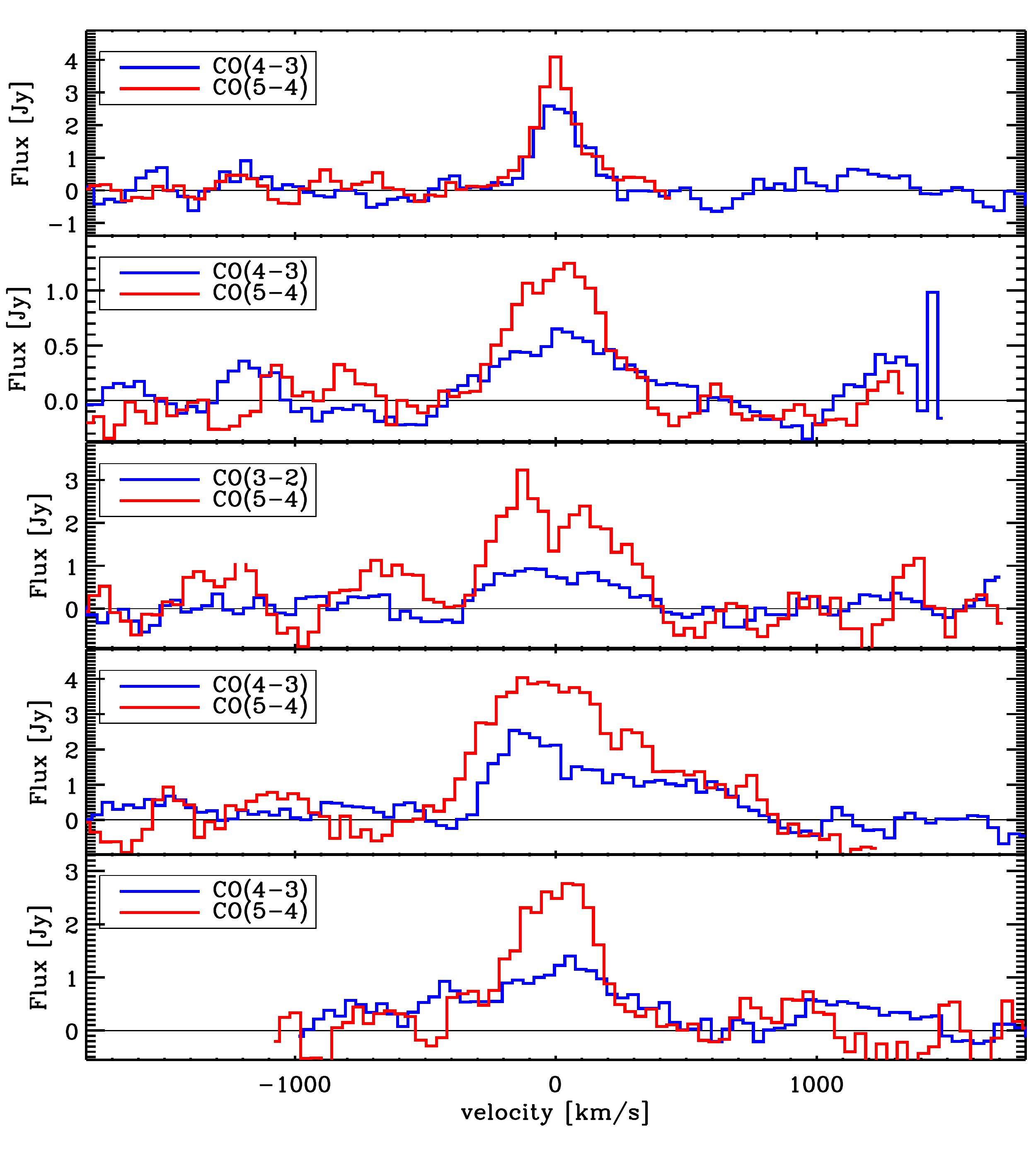}
     \caption{Mm/sub-mm spectra of the five sample galaxies in band 4 (red) and band 3 (blue), highlighting, respectively, the CO(5-4) emission and the CO(4-3) (for objects 1, 2, 4 and 5) or CO(3-2) (for object 3). The spectra, extracted in channels of 40 km/s, are expressed in Jy as a function of the velocity offset from the redshift centered on the CO(5-4) line, and are extracted from the region over which we integrate the moment 0 maps to obtain the total line fluxes. In each panel, we report the FWHM of the CO(5-4) line emission.}
             \label{Fig:spectra}%
    \end{figure}

Finally, we re-extracted the spectra from the same regions in the moments 0 maps that have signal 2$\sigma$ over the rms: the spectra are shown in Figure~\ref{Fig:spectra}, centered in velocity on the peak of the CO(5-4) line. CO(5-4) emission is well detected in all five objects, with FWHM between $\sim$150 km/s (Gal1) to $\sim$600 km/s (Gal4). The shapes of the line profiles are quite diverse: from a narrow line (Gal1) to a double peak (Gal3) to broad emission (Gal2, 4 and 5). The second line, in band 3, is also detected in all objects, although with lower significance. It is important to stress that these spectra have shapes that are very similar to the ones extracted in the first step; however, they are less noisy than those, having been extracted only from the region where the line signal is robustly detected.

\section{CO vs dust continuum vs UV}

We detect CO(5-4) emission at $\geq7\sigma$ for all five objects (and up to 13 $\sigma$ for object 4), while CO(4-3) (and CO(3-2) for gal 3) emission is detected at $\geq3\sigma$ for all objects (see Table~\ref{tab:properties_alma1}). The continuum in band 3 at $\sim$650 $\mu$m rest-frame is not detected (therefore not shown in Fig.~\ref{Fig:images}), while the continuum in band 4 at $\sim$500 $\mu$m rest-frame is always detected at $\geq4\sigma$ (therefore we included it in the SED fitting process presented in Section~2.1). We obtain line integrated fluxes  $L_{CO}$ in the range 0.4-1 Jy km/s for CO(4-3) (or CO(3-2) for object 3), and 0.6-3 Jy km/s for CO(5-4). We checked that these values are within $\pm10$\% from the values obtained by using the {\rm GAUSSFIT} or {\rm 2DFIT} tool within CASA.

In the first two panels for each galaxy in Fig.~\ref{Fig:images} we show the comparison between the positions of the two CO emission (blue and red), of the dust (green) and of the UV-optical rest-frame light (grey scale, as traced by the HST/F814W and UltraVISTA $K_s$ DR4 imaging, corresponding to $\sim$2000\AA and $\sim5000$\AA~rest-frame, respectively). In Fig.~\ref{Fig:images} we report as well the position of the spectral slit, in order to compare the regions where the CO, dust and optical spectrum originate for each galaxy.

\subsection{Gal1}
The morphology of Gal1 in the ACS/z-band (corresponding to 2000\AA~rest-frame), appears quite faint and compact, and the UltraVISTA DR4 $K_s$ image, corresponding to the 5500\AA~rest-frame, shows also a quite compact morphology, aligned within 0.1" with the ACS/z-band one (see Fig.~\ref{Fig:images}. The ALMA band 4 continuum emission is not spatially resolved, and falls ontop of the z-band/$K_s$ emission. The spectroscopic slit is well aligned with the UV-optical-ALMA emission, and Gal1 has a spectroscopic redshift $z_{opt}$=3.1120, based on the SiII$\lambda$1260.4~\AA, OI$\lambda$1303~\AA, [CII]$\lambda$1334.5~\AA, SiIV$\lambda$1393.8+SiIV$\lambda$1402.8~\AA, and on SiII$\lambda$1526.7 lines detected in the VUDS spectrum. Both CO emission lines are quite narrow, with FWHM $\sim200$ km/s (see Fig.~\ref{Fig:spectra}), and they are centered at $z_{CO}$=3.1181, meaning that the ISM is blueshifted with respect to the CO lines by $\sim400$ km/s. If we assume that the CO is a good tracer of the systemic velocity of the system, this implies that the ionized gas traced by the UV ISM lines is outflowing with velocities $\sim400$ km/s, not unusual for galaxies at these redshifts (Steidel et al. 2010; Erb et al. 2014; Marchi et al. 2019). We can conclude that this is very likely a single object, with dust that, along some line of sight, absorbs the UV light, re-emitting it in the FIR. By fitting the multi-wavelength photometry from the UV to sub-mm, we derive a stellar mass of $M_{*}=7.25\times10^{10} M_{\odot}$ and a $\delta_{MS}=1.38$ (or $\delta_{MS}=2.4$, if the FIR is used to estimate the SFR). 

The peaks of the CO(5-4), CO(4-3) and continuum in band 4 are within 0.5" of each other, and ontop the position of the UV-optical rest-frame emission, as traced by the HST/F814W and UltraVISTA $K_s$ images. Both CO lines are spatially resolved, as it can be seen in Fig.~\ref{Fig:images}. In band 4, the CO(5-4) emission has a size (deconvolved from beam) of 1.66$\pm$0.48 arcsec (major axis) $\times$ 0.74 $\pm$ 0.23 arcsec (minor axis), with a clean beam of 1.48"$\times$0.76".

\subsection{Gal2}\label{sect:gal2}
This is the object for which the multi-wavelength photometry is perturbed by a lower redshift interloper. In Fig.~\ref{Fig:images} we show the two circular apertures that we used to deblend the two components: the interloper and the actual object at $z\sim3$ are shown by the continuous and dotted line, respectively. Gal2 is very faint in the ACS/z-band (corresponding to 2000\AA~rest-frame), and shows a quite irregular and extended morphology in the optical rest-frame, probed by the UltraVISTA $K_s$ band. In particular, the object extend for $\sim$2.5" ($\sim$ 20 kpc at $z\sim3$) perpendicularly to the spectroscopic slit.  The CO(5-4) and CO(4-3) emissions are spatially unresolved, coincident with each other, but offset from the band 4 continuum emission, that is quite extended (see Fig.~\ref{Fig:images}): we could measure a size of $1.934\pm$ 0.350 arcsec (FWHM along the major axis, corresponding to 15.0$\pm$2.5 kpc) $\times 0.767\pm 0.060$ arcsec (FWHM along the minor axis, corresponding to 6.0$\pm$0.5 kpc), under a clean beam of $\sim1.47\times0.76$ arcsec. In order to investigate this further, we performed a different cleaning process only for this object, using this time a Briggs weighting scheme (Briggs 1995) with {\it robust} parameter 0.5, in order to improve the spatial resolution, without compromising too much the sensitivity. With this imaging, it appears that the band 4 continuum emission is actually bimodal, with a peak that is spatially coincident with the CO lines, and a second component that is aligned NW with respect to the first one, and lying within the spectroscopic slit. 

For this object, the CO emission is much less extended than the optical rest-frame light traced by the UltraVISTA $K_s$ imaging; moreover, the CO emission is also spatially offset from the spectroscopic slit, meaning that it originates from a region that is not probed by the optical spectroscopy. The spectroscopic redshift, re-centered on the ISM lines (based on the SiII$\lambda$1260.4~\AA, OI$\lambda$1303~\AA, SiII$\lambda$1526.7, and CIV$\lambda$1548.4~\AA lines), turns out to be $z_{opt}$=3.0494, offset by -450 km/s with respect to the CO emissions, that are quite broad (FWHM$\sim$350-500 km/s) and centered at $z_{CO}=3.0557$. It is worth noting that the ALMA tuning could reveal CO(5-4) between $z=3.0204$ and $z=3.12535$, therefore it could detect emission offset from the optical emission by $-6000$ up to $+1850$ km/s: the fact that the CO(5-4) is not detected in the region overlapping with the spectroscopic slit means that there is no (or very little, not detectable) CO in that region.

When we fit the multi-wavelength deblended photometry for Gal2 with MAGPHYS we obtain a stellar mass of $M_{*}=6.6\times10^{10} M_{\odot}$ and a $\delta_{MS}=0.69$ (or $\delta_{MS}=0.94$, if the FIR is used to estimate the SFR).

\subsection{Gal3}
The morphology of Gal3 in the UV rest-frame (traced by the ACS/i-band image) is quite irregular: a bright clump sits aside with a fainter and smoother component, separated by $\sim0.7"$. The object looks more regular in the UltraVISTA DR4 $K_s$ image, corresponding to the optical 5500\AA~rest-frame, and the emission is centered between the two UV rest-frame peaks. The continuum emission in band 4 is not spatially resolved, and lies ontop of the brightest UV peak, within $\sim0.3$" from the optical rest-frame emission.

The spectroscopic slit is well aligned with the UV-optical-ALMA emission, and the object has a spectroscopic redshift $z_{opt}=2.9342$, based on the SiII$\lambda$1260.4~\AA, OI$\lambda$1303~\AA, [CII]$\lambda$1334.5~\AA, SiII$\lambda$1526.7, and CIV$\lambda$1548.4~\AA lines, revealed in the zCOSMOS spectrum. Both CO emission lines are quite broad, with FWHM$\sim$400-450 km/s (see Fig.~\ref{Fig:spectra}), with the CO(5-4) line being clearly double peaked, with the two peaks separated by 250 km/s, possibly indicating a rotating disk, and centered at $z_{CO}=2.9348$, that implies that the ISM lines are blueshifted with respect to the CO by $45$ km/s, a difference that is not statistically significant, given the precision of the optical spectroscopy, that provides a resolution of $\sim100$ km/s. Since the spatial and spectral offsets we find are small or absent, we can conclude that this is a single object. Running MAGPHYS, we obtain for Gal3 a stellar mass of $M_{*}=3.1\times10^{10} M_{\odot}$ and a $\delta_{MS}=3.1$ (or $\delta_{MS}=4.7$, if the FIR is used to estimate the SFR). 

The CO(5-4) emission is spatially resolved, but both the CO(3-2) emission and the continuum in band 4 are not. In particular, we could measure a size of 1.92$\pm$0.55 arcsec (FWHM along the major axis, corresponding to $\sim15\pm4$ kpc) $\times$ 0.43$\pm$0.40 arcsec (FWHM along the minor axis, corresponding to $\sim3.3\pm$3 kpc), for a clean beam of 1.08$\times$0.81 arcsec. The peaks of the two CO lines and the continuum emission are within 0.2" from each other, they are offset by $\sim$0.2" from the position of the UV emission, as revealed by the ACS/i-band imaging, and are coincident with the optical emission, traced by the $K_s$ imaging.

\subsection{Gal4}
The morphology of Gal4 in the ACS/i-band, corresponding to the UV rest-frame, is double peaked, as it is the morphology in the UltraVISTA DR4 $K_s$ imaging, that matches the optical rest-frame: both peaks fall within the 1" spectroscopic slit, and are separated by $\sim0.7"$. The ALMA band 4 emission is not spatially resolved, and is centered ontop of the faintest of the two peaks revealed by the i-/$K_s$ band imaging. This might indicate that the dust is not homogeneously distributed in this object: the UV bright clump indicate a region free of dust, from which the UV photons are free to escape; the region where the ALMA band 4 continuum is emitted is, on the other hand, rich of dust that absorbs the UV photons and re-emit them in the FIR/sub-mm. The spectroscopic slit is well aligned with the UV-optical-ALMA emission, and the object has a spectroscopic redshift $z_{opt}=3.4388$, based in this case on the peak of the Lyman$\alpha$ line (the ISM absorption lines are in this case very faint). 

Both CO emission lines are very broad, with FWHM $\sim500$ km/s (see Fig.~\ref{Fig:spectra}), have an asymmetric shape, with the lines being skewed for high velocities, and centered at $z_{CO}=3.4315$. This implies that the Lyman$\alpha$ emission is redshifted with respect to the CO by $\sim500$ km/s, not unusual for galaxies at these redshifts (Steidel et al. 2010; Erb et al. 2014; Marchi et al. 2019). We can conclude, again, that, since the spatial offsets between UV-optical and FIR/sub-mm are small, and there is no evidence of other components from the optical and ALMA spectra, Gal4 is a single object. Running MAGPYS on the multi-wavelength photometry provides for Gal4 a stellar mass of $M_{*}=6.6\times10^{10} M_{\odot}$ and a $\delta_{MS}=3$ (or $\delta_{MS}=10$, if the FIR is used to estimate the SFR). 

The CO(5-4) emission is spatially resolved, and the CO(4-3) emission is marginally resolved, while the continuum emission in band 4 is unresolved. We could measure a size of 1.315$\pm$0.156 arcsec (FWHM along the major axis, corresponding to 10$\pm$1.15 kpc) $\times$ 0.606$\pm$0.094 arcsec (FWHM along the minor axis, corresponding to 4.5$\pm$0.7 kpc), under a clean beam of 0.73$\times$0.60 arcsec, for the CO(5-4) emission. As it can be seen from Fig.~\ref{Fig:images}, the CO(5-4), CO(4-3) and band 4 continuum all sit ontop of each other, and the emissions are coincident with the center of the  map, where the multi-wavelength photometry  and the spectrum were extracted. The ALMA continuum and CO emissions are offset by $\sim0.8"$ from the brightest clump detected in the UV-optical rest-frame, probed by the HST i-band and UltraVISTA $K_s$ images.

\subsection{Gal5}
The morphology of Gal5 is quite irregular in the ACS/i-band imaging, tracing the UV rest-frame, with a bright clump and a faint feature, that could resemble a tidal tail, separated by 0.5". Gal 5 is, on the other hand, more regular in the UltraVISTA DR4 $K_s$ imaging, that traces the optical rest-frame, with the emission lying right ontop of the UV light. The continuum in ALMA band 4 is not spatially resolved, and is well aligned with the UV and optical rest-frame emission. The spectroscopic slit is well aligned with the UV-optical-ALMA light, and the object has a spectroscopic redshift $z_{opt}=3.3428$, based on Ly$\alpha\lambda$1215.7~\AA in absorption, SiII$\lambda$1260.4~\AA, SiIV$\lambda$1393.8+SiIV$\lambda$1402.8~\AA, SiII$\lambda$1526.7, and CIV$\lambda$1549.5~\AA lines. Both CO emission lines are quite broad,  with FWHM$\sim$400-700 km/s (see Fig.~\ref{Fig:spectra}), and centered at $z_{CO}=3.3411$, implying an offset between ISM lines and CO of +100 km/s, not statistically significant due to the quite low spectral resolution provided by the VUDS spectroscopy.
Again, this indicates that this is very likely a single object. Running MAGPHYS on the multi-wavelength photometry, we obtain a stellar mass of $M_{*}=3.7\times10^{10} M_{\odot}$ and a $\delta_{MS}=3.3$ (or $\delta_{MS}=4.9$, if the FIR is used to estimate the SFR).

The CO(5-4) emission is spatially resolved, while the CO(4-3) emission is not. We could measure a size of 0.998$\pm$0.143 arcsec (FWHM along the major axis, corresponding to 7.5$\pm$1.1 kpc) $\times$ 0.566$\pm$0.101 arcsec (FWHM along the minor axis, corresponding to 4.2$\pm$0.75 kpc), under a clean beam of 0.62$\times$0.55 arcsec, for the CO(5-4) emission. As it can be seen from Fig.~\ref{Fig:images}, the CO(5-4), CO(4-3) and band 4 continuum all sit ontop of each other, and  they are coincident with $K_s$ emission, that is at the center of the map.

\begin{table*}[ht]
    \centering
\begin{tabular}{c c c c c c c}%{c|c|c|c|c|c}
\hline\hline \noalign{\smallskip}
      & $z_{CO}$ & $L_{CO(4-3)}\Delta v$ & $L_{CO(5-4)}\Delta v$ & F(band4) & $\sigma_e^a$ & $D^b$\\
            &  &  [Jy km s$^{-1}$] & [Jy km s$^{-1}$] & [$\mu$Jy] & [Km/s] & [kpc]\\
  \noalign{\smallskip} \hline
\noalign{\smallskip}

 Gal1 & 3.1181 & 0.58$\pm$0.06 & 0.71$\pm$0.05 & 132$\pm$32 & 83 & 14.48$\pm$1.52\\
 Gal2 & 3.0557 & 0.50$\pm$0.11 & 0.52$\pm$0.11 & 212$\pm$29 & 161 & 8.28$\pm$1.30\\
 Gal3 & 2.9348 & 0.47$\pm$0.16$^a$ &  1.28$\pm$0.16 & 237$\pm$43 & 193 & 10.08$\pm$1.16\\
 Gal4 & 3.4315 & 1.10$\pm$0.12 &  2.90$\pm$0.08 & 111$\pm$22 & 279 & 9.58$\pm$0.59\\
 Gal5 & 3.3411 & 0.73$\pm$0.13 &  1.09$\pm$0.06 & 156$\pm$23 & 158 & 8.35$\pm$0.52\\
\noalign{\smallskip} \hline
\noalign{\smallskip}
 
     \end{tabular}
    \tablenotetext{a}{Standard deviation, or dispersion, of the gaussian that fits the line profile; the Full Width Half Maximum can be obtained as FWHM=$\sigma_e\times$2.35.}
    \tablenotetext{b}{Deconvolved from the instrumental beam}
   \caption{Measurements on ALMA data for the sample} 
    \label{tab:properties_alma1}
\end{table*} 

\begin{table*}[ht]
    \centering
\begin{tabular}{c c c c c c c c c}%{c|c|c|c|c|c}
\hline\hline \noalign{\smallskip}
      & $L'_{CO(5-4)}$ & $M_{gas,CO}$ & $M_{gas,band 4}$ & $\mu_{gas,CO}^a$ & $\mu_{gas,band4}^a$ & $t_{depl,CO}$ & $t_{depl,band4}$ & $M_{dyn}$\\
      & [$10^{10}$ K km $s^{-1} pc^2$] & [$10^{10} M_{\odot}$] & [$10^{10} M_{\odot}$] &  &  & [Gyr] & [Gyr] & [$10^{11} M_{\odot}$]\\
  \noalign{\smallskip} \hline
\noalign{\smallskip}

 Gal1 & 1.20$\pm$0.20 & 13.52$\pm$1.35  & 11.87$\pm$2.87 & 0.66 & 0.63 & 0.44 & 0.39 & 0.47$\pm$0.06\\
 Gal2 & 8.39$\pm$0.18 & 9.48$\pm$2.01   & 19.21$\pm$2.64 & 0.59 & 0.74   & 0.64 & 1.29 & 1.09$\pm$0.19\\
 Gal3 & 1.95$\pm$0.24 & 10.29$\pm$1.28  & 19.00$\pm$3.41 & 0.77 & 0.86 & 0.31 & 0.56 & 1.80$\pm$0.23\\
 Gal4 & 5.74$\pm$0.57 & 21.79$\pm$2.18  &  8.34$\pm$1.65 & 0.77 & 0.56 & 0.35 & 0.13 & 3.57$\pm$0.29\\
 Gal5 & 2.05$\pm$0.20 & 10.84$\pm$1.08  & 11.68$\pm$1.76 & 0.63 & 0.65 & 0.15 & 0.16 & 1.47$\pm$0.14\\
\noalign{\smallskip} \hline
\noalign{\smallskip}
 
     \end{tabular}
     \tablenotetext{a}{Defined as $M_{gas}/(M_{gas}+M_*)$}
   \caption{Derived properties of the sample from ALMA data} 
    \label{tab:properties_alma2}
\end{table*} 

\section{CO luminosity and SLED slope}
We obtain CO(5-4) luminosities from the fluxes by applying the following equation from Solomon, Downes ~\&~ Radford~(1992):
\begin{equation}
\begin{split}
\\
&  \hspace{-1cm}L'_{CO(5-4)}=\\   
  & \hspace{-.5cm}=3.25\times10^7\times S_{CO(5-4)}\Delta v \times\\
  & \hspace{1cm}\times\frac{D_L^2}{(1+z)^3\nu_{obs}^2} K km/s~pc^2,
    \end{split}
\end{equation}

where $S_{CO(5-4)}\Delta v$ is the velocity integrated line flux, $D_L$ is the luminosity distance, $\nu_{obs}$ is the observed frequency of the emission, and $z$ is the redshift.

\begin{figure}[!ht]
   \centering
  \includegraphics[width=\columnwidth]{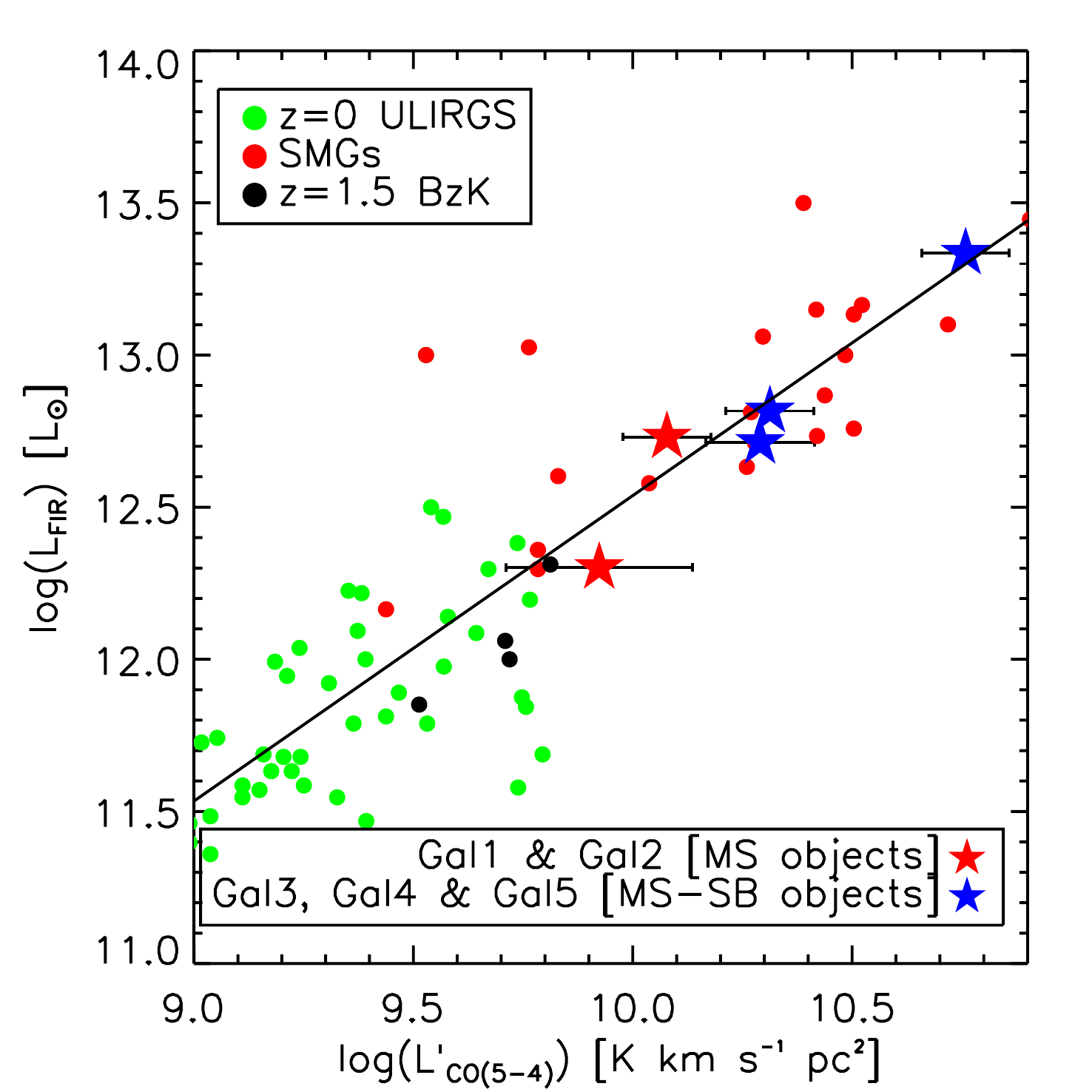}
     \caption{Far-infrared luminosity, obtained with MAGPHYS from fitting the spectral energy distribution, as a function of CO(5-4) luminosity $L'_{CO(5-4)}$, for the five galaxies in our sample (red stars: normal MS galaxies; blue stars: galaxies at the boundary to starbursts) together with literature data points (compilation by Daddi et al. (2015): green and red circles are z=0 ULIRGs and SMGs, respectively, from Magdis et al. (2012) and Carilli~\&~Walter (2013); black circles are BzK galaxies at $z\sim1.5$ by Daddi et al. 2015). The diagonal line is the linear correlation proposed by Daddi et al. (2015).
     }
\label{Fig:COlum_FIR}
    \end{figure}

We report in Table~\ref{tab:properties_alma2} the CO(5-4) luminosities $L'_{CO}$, and in Figure~\ref{Fig:COlum_FIR} we plot them against $L_{FIR}$, in comparison with data from literature, including local SMGs and ULIRGs by Magdis et al. (2012), Carilli~\&~Walter(2013) and BzK galaxies at $z\sim1.5$ by Daddi et al. (2015). Our objects are at the high-end of the distribution of points, and they distribute quite well around the linear correlation proposed by Daddi et al. (2015). It is interesting to note that the two galaxies that lie closer to the average MS at $z\sim3$, Gal1 and Gal2, have also the smallest $L'_{CO(5-4)}$. Our measurements therefore confirm that the correlation between the CO(5-4) line and FIR luminosities, observed at $z\sim0$ and $z\sim1.5$, is still in place at $z>3$ for MS galaxies.

\begin{figure*}[!ht]
   \centering
  \includegraphics[width=\textwidth]{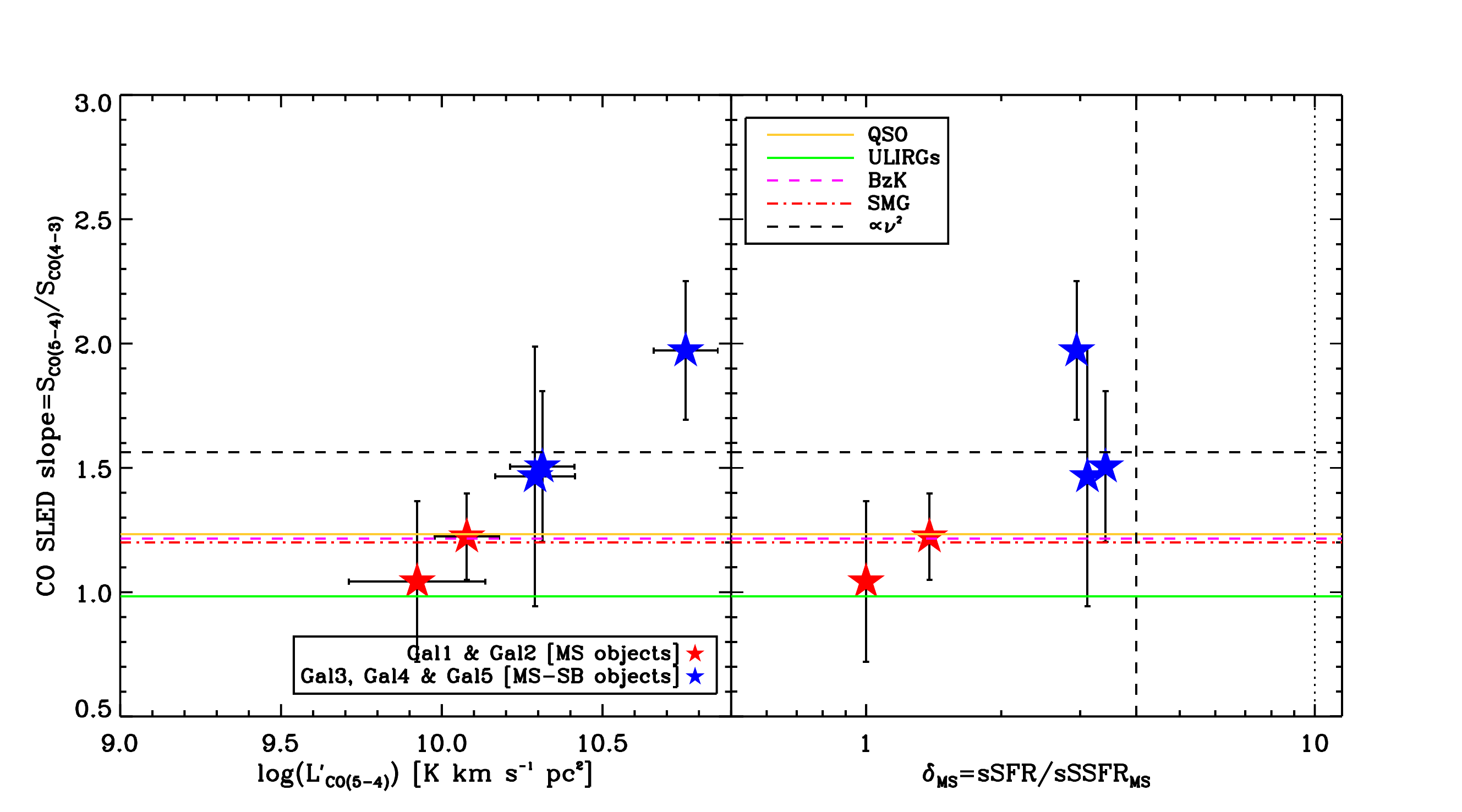}
     \caption{The CO Spectral Line Energy Distribution (SLED) slope between CO(5-4) and CO(4-3), defined as $S_{CO(5-4)}/S_{CO(4-3)}$, as a function of the CO(5-4) luminosity $L'_{CO(5-4)}$ (left panel) and of $\delta_{MS}$ (right panel: for clarity here we show the value obtained using the SFR derived by {\it MAGPHYS} only, but the result does not change if we use the SFR derived from the FIR), for the five galaxies in the sample. The colored horizontal lines show the average SLED slope for different classes of objects: continuous yellow for the median of the QSOs from Carilli~\&~Walter (their Figure 4); dot-dashed red for SMGs (Bothwell et al. 2013); dashed magenta for BzK (Daddi et al. 2015); and green for local ULIRGs (Papadopoulos et al. 2012). The dashed horizontal line shows the slope for the case of constant brightness temperature on the Rayleigh-Jeans scale, i.e., $S\propto\nu^2$. The vertical lines in the left panel highlight the locii 4x and 10x above the MS. The two galaxies that are closer to the MS (Gal1~\&~Gal2) are shown with red stars, while the three that are at the boundary between MS and starbursts are shown in blue. Error bars are estimated by propagating the errors on the individual fluxes to the ratio.
     }
\label{Fig:SLED_slope}%
    \end{figure*}

    The shape of the CO Spectral Line Energy Distribution (CO SLED) can be used to investigate the nature of galaxies by constraining properties of the energy source that excites the ISM (Carilli~\&~Walter~2013). In Figure~\ref{Fig:SLED_slope} we compare the CO SLED slope between CO(5-4) and CO(4-3), defined as $SL_{5/4}=S_{CO(5-4)}/S_{CO(4-3)}$, for the five galaxies in this sample, as a function of  $L'_{CO(5-4)}$ and $\delta_{MS}$, with values for various types of galaxies in the literature.

Since for Gal3 we targeted and observed CO(3-2) instead of CO(4-3), for that object we linearly interpolate the CO(5-4) and CO(3-2) fluxes to obtain $S_{CO(4-3)}$. By looking at Fig.~\ref{Fig:SLED_slope} it is clear that Gal1 and Gal2, the objects that lie closer to the MS, have CO SLED slopes that are compatible with those of various classes of star-forming galaxies, as local ULIRGs (Papadopulos et al. 2012), BzK at $z\sim1.5$ (Daddi et al. 2015), SMGs (Bothwell et al. 2013) and to QSO as well (Carilli~\&~Walter~2013); these two objects are also the ones with the lowest $L'_{CO(5-4)}$ luminosities. On the other hand, Gal3, 4 and 5, those that lie close to the boundary between the MS and  starbursts (SBs), and that have the highest $L'_{CO(5-4)}$ luminosities, have much steeper CO SLED slopes, in one case even in excess of those expected for a constant brightness temperature on the Rayleigh-Jeans scale, i.e. $S\propto\nu^2$. It is important to stress that none of the five galaxies in this sample shows signs of the presence of an AGN, at any wavelength. A Spearman correlation test ($r_S$=0.9) confirms that a positive correlation exists between the $SL_{5/4}$ parameter and $L'_{CO(5-4)}$. We checked that the $SL_{5/4}$ parameter does not correlate with other parameters, such as gas or stellar mass, continuum luminosity, or source size, but shows a similar correlation with sSFR (right panel of Figure~\ref{Fig:SLED_slope}). This indicates that $SL_{5/4}$ as well correlates with $L_{FIR}$ and therefore star-formation. It is also significant that we do see a correlation between distance from the MS and the dust temperature fitted by MAGPHYS: Gal4~\&~5, that are more offset from the MS and have higher $SL_{5/4}$ and $L'_{CO(5-4)}$, have higher dust temperatures, in excess of 50 K (see Table~\ref{tab:properties_photo}), while the other 3 galaxies have temperatures around 35 K (except for Gal1, for which however the temperature is not very well constrained).

  \begin{figure}[ht]
   \centering
  \includegraphics[width=\columnwidth]{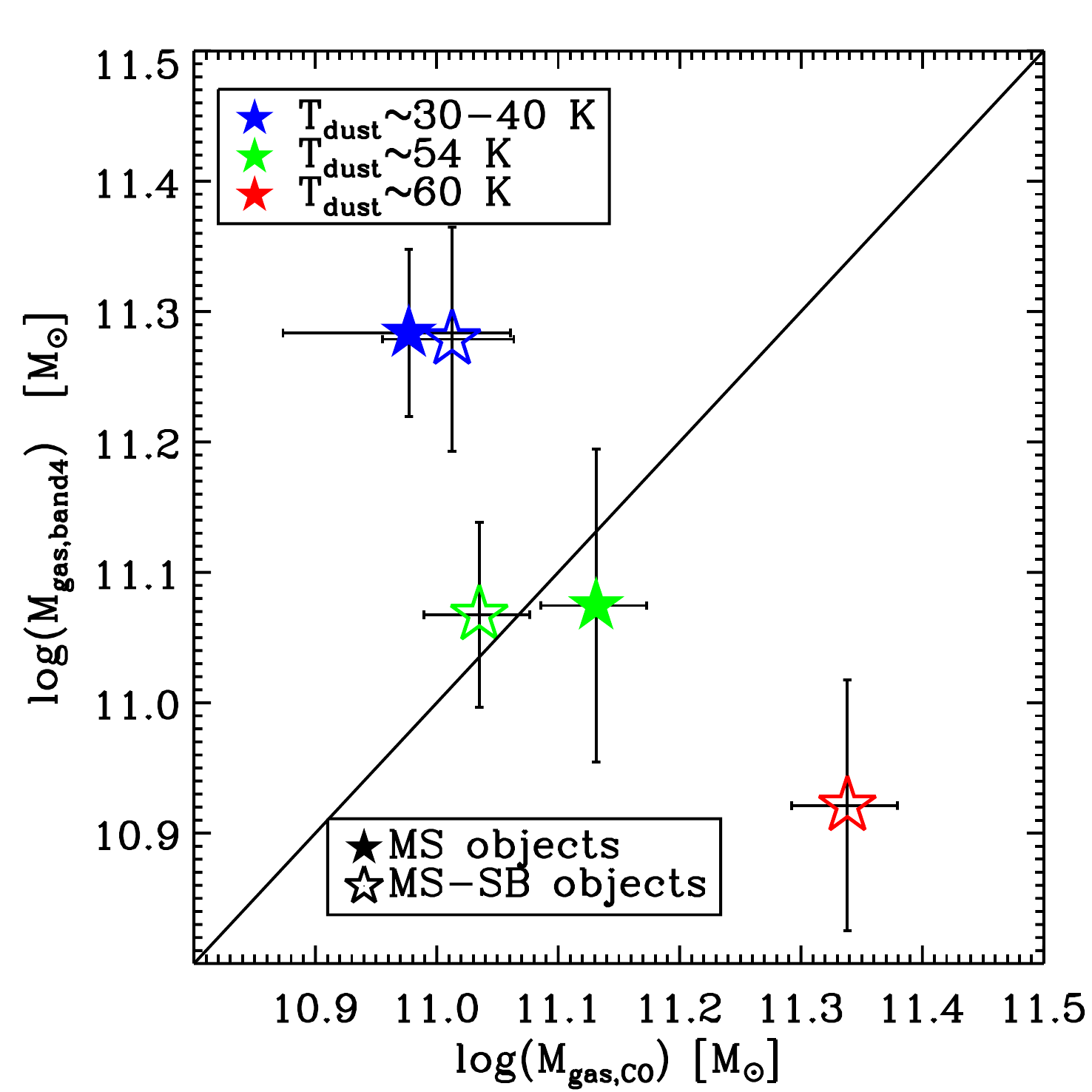}
     \caption{Comparison of the molecular gas mass estimated via the CO and dust continuum. Different colors indicate different dust temperatures, as estimated by MAGPHYS; filled symbols show the two galaxies that are closer to the MS, while the empty ones show the three galaxies that lie close to the boundary with the SB sequence.
     }
\label{Fig:CO_vs_band4}%
    \end{figure}

  \begin{figure}[ht]
   \centering
  \includegraphics[width=\columnwidth]{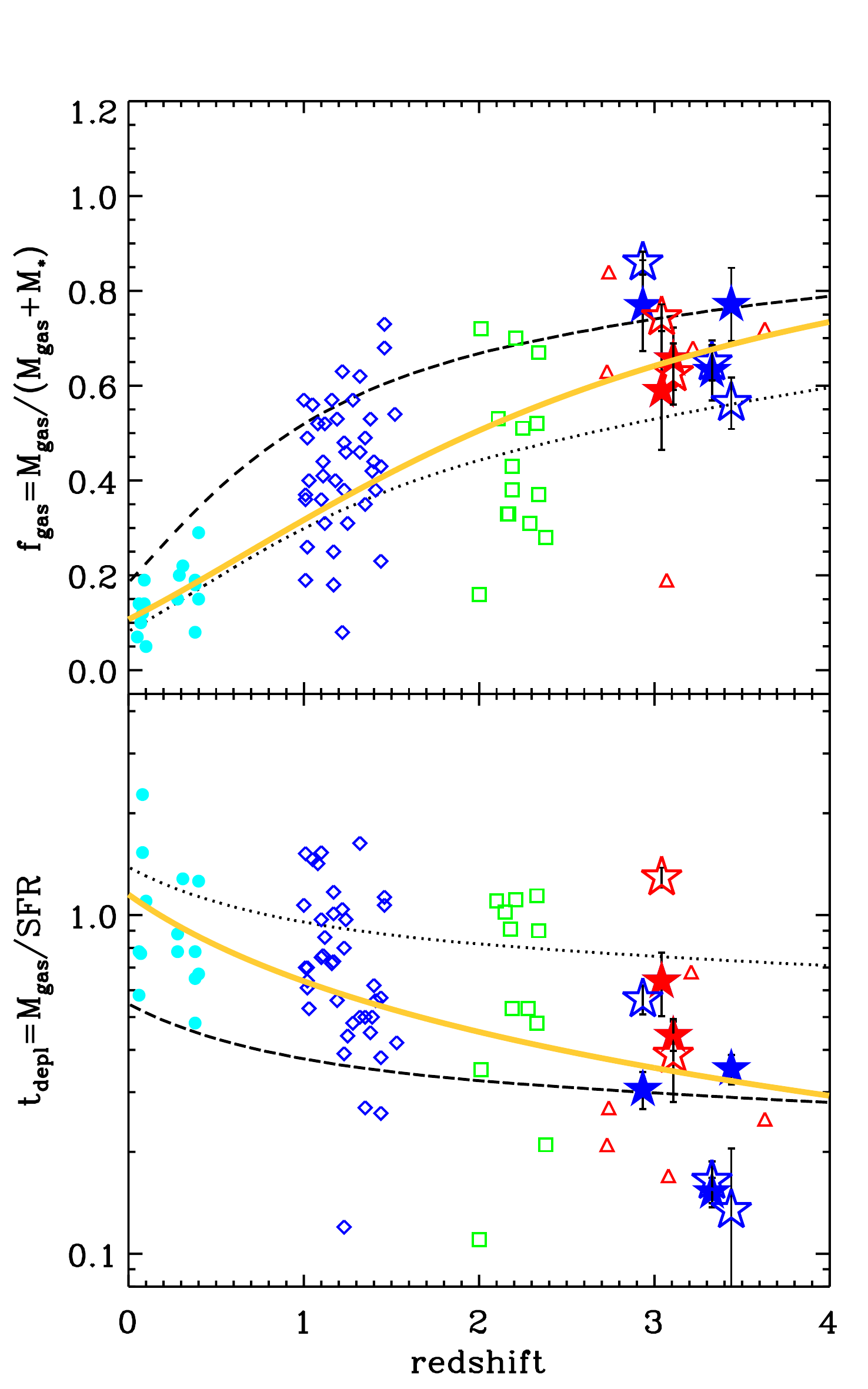}
     \caption{As a function of redshift, gas fraction (upper panel) and gas depletion time (lower panel) for the five galaxies in our sample. Red stars indicate the two galaxies closer to the MS, the blue stars indicate the objects at the boundary with the SBs, respectively, and filled and empty stars indicate measurements with gas masses obtained through CO and band 4 continuum, respectively. We also show measurements from literature by Dessauges-Zavadsky et al. (2015, 2017), for MS galaxies, divided in redshift bins: cyan at $z<0.5$, blue at $1<z<1.5$, green at $2<z<2.5$, and red at $2.8<z<3.5$. The dotted and dashed indicate the predictions of the 2-SFM model by (Sargent et al. 2014) for a galaxy with stellar mass $4\times10^{10} M_{\odot}$ lying exactly on the average MS or a factor of 3 above the MS, respectively. The solid orange line indicates the fits to the data proposed by Dessauges-Zavadsky et al. (2017): $f_{gas}=1/(1 + (0.12\times(1 + z)^{1.95})^{-1})$ and $t_{depl}=1.15\times(1 + z)^{-0.85}$.
     }
\label{Fig:tdepl}%
    \end{figure}
    
\section{Gas masses from CO and band 4 continuum}
As we discussed in Section~1, both CO lines and the dust continuum in the Rayleigh-Jeans regime can be used as tracers for the molecular gas (e.g. Scoville et al. 2016; Genzel et al. 2015). In this section we present and compare the gas masses determined using the two methods.

In order to obtain an estimate of the molecular gas mass from our CO measurements, we have to make two assumptions: the first is about the flux ratio between the CO(5-4) and CO(1-0) transitions; the second assumption is about the $\alpha_{CO}$ conversion factor between the CO(1-0) luminosity and the molecular gas mass (for a review see Bolatto, Wolfire and Leroy 2013). Although extrapolating the CO(1-0) luminosity from the CO(5-4) one can be tricky (see Carilli~\&~Walter 2013, Daddi et al. 2015), in our case we have the advantage that we detect a second lower transition for all objects, that we use to obtain at least a first guess on the overall shape of the CO SLED. By using the values presented in Section 4, and in particular in Figure~\ref{Fig:SLED_slope}, we calibrate the flux ratios between CO(5-4) and CO(1-0) for the five galaxies in our sample: for Gal4, the one with the most extreme slope, we use a $L'_{CO(5-4)}/L'_{CO(1-0)}$=25; for Gal3~\&~Gal4, both of which have intermediate slopes, we use $L'_{CO(5-4)}/L'_{CO(1-0)}$=18; for Gal1~\&~Gal2 we use $L'_{CO(5-4)}/L'_{CO(1-0)}$=8.4. These values are in line with the ones measured for objects observed in literature, that have similar $SL_{5/4}$ slopes (see Figure~4 in Carilli~\&~Walter~(2013) for a compilation): M82 for example has $L'_{CO(5-4)}/L'_{CO(1-0)}\sim$20, and BzKs have $L'_{CO(5-4)}/L'_{CO(1-0)}\sim$6.5, not far from the factors that we used here.

For the $\alpha_{CO}$ conversion factor we instead decided to use a single value for all galaxies, and we chose the metallicity dependent $\alpha_{CO}$ factor suggested by Tacconi et al. (2018) for galaxies on the MS: applying their equations (2) and (4) at the median redshift of our sample, for a stellar of mass $10^{10.5} M_{\odot}$, we obtain $\alpha_{CO}=3.8~M_{\odot} (K~km~s^{-1}~pc^{-2})^{-1}$.

{In order to obtain an estimate of the gas mass from the dust continuum in the Rayleigh-Jeans regime, we apply the method by Groves et al. (2015), that provides calibrations of the gas mass - dust luminosity relation at different wavelengths. Our band 4 observations probe the continuum very close to the 500$\mu$m rest-frame, therefore we use the calibration at that wavelength. However, since the tuning of ALMA observations is different for each object, and in addition they are at slightly different redshifts, the band 4 observations correspond to slightly different rest-frame wavelengths, between $\sim480$ and $\sim550$ $\mu$m rest-frame: in order to account for that, we correct the observed fluxes by extrapolating along the $F\propto \lambda^{-4}$ Rayleigh-Jeans law, from the observed wavelength to exactly 500 $\mu$m; the corrections are between 0.7 and 1.15. We then convert fluxes to luminosities, and use the conversion in Groves calibrated for galaxies with $M/M_{*}>10^9$ to obtain gas masses.}

We report the molecular gas masses that we obtain with these two procedures in Table~\ref{tab:properties_alma2}: they range from $\sim9\times10^{10}$ to $2.2\times10^{11}$ $M_{\odot}$, implying very high molecular gas fractions $\mu_{gas}=M_{gas}/(M_{gas}+M_*)$, between 65 and 80\%. We also estimate the gas depletion times $t_{depl}=M_{gas}/SFR$, using the SFR estimated by MAGPHYS: we obtain values in the range $0.2<t_{depl}<1$ Gyr. Both $f_{gas}$ and $t_{depl}$ are given in Table~\ref{tab:properties_alma2}. In Figure~\ref{Fig:CO_vs_band4} we compare the molecular gas estimated via the CO line and dust continuum: only for two out of five objects in the sample the two estimates agree, within the errors. It is interesting to note that these two objects have very similar dust temperatures as estimated by MAGPHYS, around $T_{dust}\sim45$ K; the two objects for which the dust is colder (with $T_{dust}\sim35-40$ K) are those for which the dust based gas mass is larger than the CO one; and conversely, the object which warmer dust temperature ($T_{dust}\sim60$ K) is the one for which the dust based gas mass is smaller than the CO one. In the same figure, we use different symbols to see if this trend could be driven by the distance from the main sequence $\delta_{MS}$, but it does not seem to be the case: the three galaxies that lie closer to the MS-SB separation are equally above, on, and below the diagonal in Figure~\ref{Fig:CO_vs_band4}. Summarizing, this could imply that the calibration to obtain gas mass from the 500 $\mu$m rest-frame flux gives estimates that are closer to the ones from the CO for galaxies with dust temperatures around 35-40 K; for galaxies with warmer (colder) dust temperature, the dust emission would be shifted towards lower (larger) wavelengths, the flux in the Rayleigh-Jeans regime would decrease (increase), and one would get a smaller (larger) flux, and therefore a smaller (larger) gas mass.

In Figure~\ref{Fig:tdepl} we show $\mu_{gas}$ and $t_{depl}$ as a function of redshift, together with the values for MS galaxies by Dessauges-Zavadsky~et~al.~(2015, 2017), who compiled a list of star-forming galaxies with $0.3<sSFR/sSFR_{MS}<3$ for which a determination of the molecular gas mass was available based on CO line measurements. Our data almost double the number of molecular gas measurements at $3 < z <3.4$, in a range where most of the available measurements are for lensed systems (four out of five galaxies in Dessauges-Zavadsky et al. 2017). Our derived gas fractions are all above the best fit curve $f_{gas}=1/(1 + (0.12\times(1 + z)^{1.95})^{-1})$ by Dessauges-Zavadsky~et~al.~(2017), but probably their best fit is somewhat biased low at $z>2.8$ by the lone outlier that has a very low gas fraction. Otherwise, our objects lie in the same region at $0.6<f_{gas}<0.8$ as the other galaxies by Dessauges-Zavadsky~et~al.~(2017). Our measurements then confirm the increase in gas fractions of MS galaxies at $z>2$ that was hinted by previous observations. The depletion times for our five galaxies are also comparable with the measurements from Dessauges-Zavadsky~et~al.~(2017) in the same redshift range, and distribute quite well around the best fit curve $t_{depl}=1.15 \times (1 + z)^{-0.85}$ by Dessauges-Zavadsky~et~al.~(2017), confirming the observed decrease of $t_{depl}$ with redshift.

In Figure \ref{Fig:tdepl} we report also the predictions of the 2-SFM model by Sargent et al. (2014), for a galaxy of $4\times10^{10} M_{\odot}$ (average stellar mass in our sample), lying exactly on the MS, or 3 times above it (the median $\delta_{MS}$ of our sample). These predictions are based on the combination of two scaling relations: (i) the evolution of the star-forming main sequence, and (ii) the integrated Schmidt-Kennicutt relation (assuming it does not evolve with redshift). In this framework, these curves provide a zero-level physical interpretation of the evolutionary trends and the position galaxies in both panels. It is interesting that all five galaxies in the sample have quite similar gas fractions (also the gas masses range a span of 0.3 dex only), and they all lie in the region of the $f_{gas}$ vs z plane comprised between the prediction for the average MS and 3 times above it: this indicate that they have $f_{gas}$ fractions in line with the prediction of the model for MS galaxies at that redshift. On the other hand, the five galaxies have more spread out values of the gas depletion timescales: the two galaxies that lie closer to the average MS (Gal1~\&~Gal2) have $t_{depl}$ quite longer than the other three galaxies, and they lie closer to the prediction for average MS galaxies; the other three have smaller $t_{depl}$, closer (and beyond) the prediction for galaxies 3 times above the average MS.

\section{Dynamical masses and $\alpha_{CO}$}
\begin{figure}[!ht]
   \centering
  \includegraphics[width=\columnwidth]{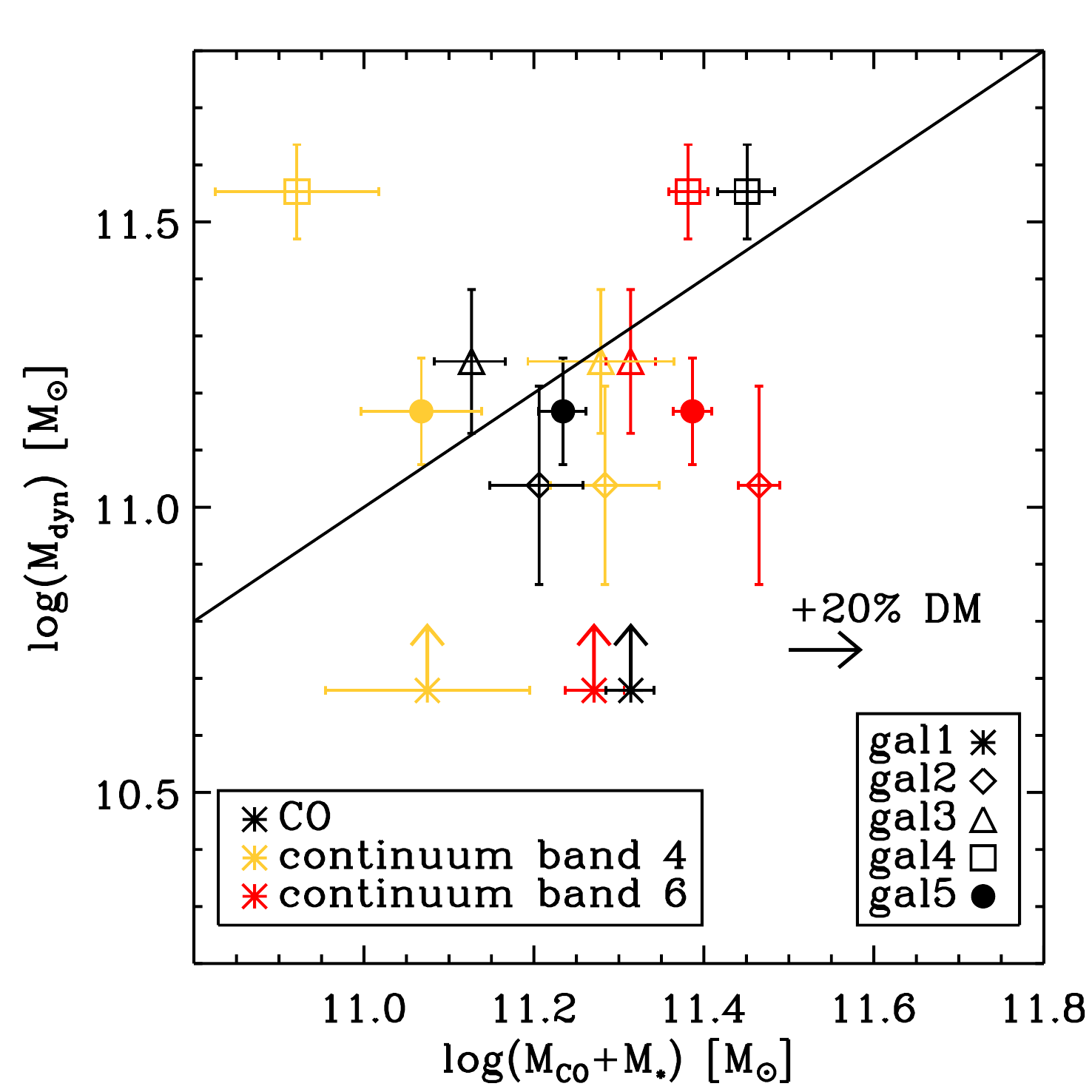}
     \caption{Comparison between dynamical mass and the sum of gas mass and stellar mass, for different methods to constrain the gas mass: black, red and orange symbols indicate the values based on CO, continuum in band 4 at 240 GHz, and continuum in band 4 at 140 GHz, respectively. The gas masses based on CO are obtained using $\alpha_{CO}=3.8~M_{\odot} (K~km~s^{-1}~pc^{-2})^{-1}$, the value suggested by Tacconi et al. (2018) at the median redshift of our sample for galaxies of similar stellar mass. The dynamical mass obtained for Gal1 is indicated as a lower limit, as probably the true inclination is much smaller than the one we obtained from the ALMA images.
     The right pointing arrow above the legend shows the total mass increase should 20\% of DM be added.}
\label{Fig:dynamical}%
    \end{figure}

In order to obtain an estimate of the dynamical masses of our objects we apply the method outlined by Wang~et~al.~(2013) and applied among others by Capak et al. (2015):  $M_{dyn} = 1.16\times10^5 V_{cir}^2 D$, where $V_{cir}$ is the circular velocity in $km/s$ and D is the diameter in kpc. The circular velocity is assumed to be $V_{cir}$= 1.763 $\sigma_{CO(5-4)}/sin(i)$, where $\sigma_{CO(5-4)}$ is the velocity dispersion, and $i$ is the disk inclination angle. We estimated the inclination angle from the ALMA images, presented in Fig.~\ref{Fig:images}, as $i=\cos^{-1}(b/a)$: the axial ratio that we used is the one calculated on the deconvolved sizes, and the inclinations that we obtain range between $i=45^{\circ}$ and $i=60^{\circ}$. For Gal2, for which the emission in both CO lines is unresolved, we assumed $i=57^{\circ})$, the most probable value in case of random orientation. This method provides a good first guess of the dynamical mass, but it suffers from the quite large uncertainties on the size and axial ratio, that can not be constrained very robustly, due to the limited spatial resolution and the elongation of the beam for some of our ALMA observations. We report the dynamical masses in Tab.~\ref{tab:properties_alma2}; moreover, in Figure~\ref{Fig:dynamical} we compare $M_{dyn}$ with the sum of stellar mass $M_*$ and gas mass $M_{gas}$, for gas masses based on different techniques (converting the CO line luminosity, and from the continuum in band 4~\&~6). 

It is interesting to note that the measurements from CO scatter quite nicely around the 1:1 relation, apart for Gal1, that has a very small velocity dispersion $\sigma_{CO(5-4)}$, leading to a very small $M_{dyn}$ (it is quite possible that the axial ratio for this object is poorly estimated, as the beam of the band4 observations is quite elongated and could have led to an overestimation of $i$; therefore we indicate its dynamical mass as a lower limit). We did not include any dark matter (DM) in these calculations, to simplify the interpretation: however, a DM fraction of 20\% (similar to the one observed by Genzel et al. (2017) and Lang et al. (2017) in galaxies of similar stellar mass at $z\sim2$) would only slightly increase the total mass ( see the arrow in Figure~\ref{Fig:dynamical}). This implies that the mass bound in stars, plus the mass in the cold gas phase, (also in the case we added 20\% of dark matter), {is comparable with } the dynamical mass in these galaxies. This suggests that all the assumptions we made in constraining the dynamical mass and the molecular gas mass are at least reasonable. In particular, it is reasonable to assume a common value for $\alpha_{CO}$ for all galaxies, with a value that is typical of MS galaxies: for example, having assumed an $\alpha_{CO}$ value more typical of starburst galaxies, $\alpha_{CO}=0.8~M_{\odot} (K~km~s^{-1}~pc^{-2})^{-1}$, would have led to total masses more than 3 times smaller than the dynamical ones: with the velocity dispersions that we observe, the only way to decrease the dynamical masses by a similar amount to reconcile them with the total gas+star masses would have be then to assume that our galaxies are seen almost edge-on, an assumption that is not at all supported by the observations.

This comparison provides at least a first constraint on the $\alpha_{CO}$ parameter for MS galaxies at $3<z<3.5$: the value that we assumed using the recipe provided by Tacconi et al. (2018), based mainly on continuum derived molecular gas masses, turns out to be the one that is needed to obtain a total mass that matches the dynamical mass. Even assuming that the inclination angle is overestimated for all galaxies, that would mean that the true dynamical masses would be even larger, and we would need a larger $\alpha_{CO}$ factor to recover the total dynamical mass, or a substantial amount of dark matter.

As a check, in Figure~\ref{Fig:dynamical} we also compared the total dynamical mass to the sum of stellar and gas mass, if the continuum in the Rayleigh-Jeans regime is used as a tracer of the molecular gas mass (similar to Scoville et al. 2014; 2016; Groves et al. 2015). In particular, we used the gas masses published by Schinnerer et al. (2016), based on the continuum observed at 240 GHz in band 6 for these same five objects, and we calculated also the masses using the continuum in band 4 at 140 GHz, already presented in Section 5. The difference between CO and continuum based gas masses are already presented in Figure~\ref{Fig:CO_vs_band4}: galaxies with colder (warmer) dust tend to have larger (smaller) dust based gas masses. It can be seen that the gas masses based on CO are the ones that give the best agreement with the dynamical masses; the continuum in band 4 seems to provide reasonable estimates for galaxies 2,3 and 5, but not for Gal4, that is the galaxy with the warmest dust, and therefore the smallest dust continuum based gas mass; the continuum in band 6 provides a good estimate for Gal4, but seems to give too large molecular gas masses for the other objects. It is important to stress that this sample is the first at $z > 3$ for which the different methods to constrain the gas mass can be compared: although the sample is quite small, it is clear that the three techniques give result that are not that far from each other.

\section{Summary}
In this paper we presented ALMA observations of two CO mid-high transitions for a sample of five star-forming galaxies across the main sequence at $3<z<3.5$. Two of them lie very close to the average MS, as constrained by Schreiber et al. (2015); the remaining three lie at the boundary between MS and SBs. For all objects we detected the CO(5-4) transition at more than 7$\sigma$, and we detected a lower CO transition (CO(4-3) for 4 objects and CO(3-2) for the remaining one) at more than 3$\sigma$. In addition, we also detected the continuum in band 4, at $\sim140 GHz$, corresponding to 500$\mu$m rest-frame, at better than 4$\sigma$. These five new detections double the number of star-forming galaxies with multiple CO detections in a region of the $sSFR/sSFR_{MS}$ vs redshift plane that is so far scarcely populated, and where most of the objects studied so far are lenses (see Fig.~\ref{Fig:deltams}).

Our main findings are:
\begin{itemize}
    \item From a multi-wavelength spectro-photometric analysis, the galaxies in our sample have similar properties to normal star-forming galaxies at z$>$3: they span the upper half of the star-forming MS a region that is populated by the less star-forming among the SMGs and galaxies selected through the BzK technique (Fig.~\ref{Fig:deltams}). They display modest but significant spatial offsets between the position of the UV rest-frame emission and the dust component, indicating large amount of dust that is not homogeneously distributed in the galaxies, blocking UV radiation along some line of sights while letting it through others. In general, the stellar component is more aligned with the dust continuum, as it is expected since optical light is less attenuated than UV light. The CO emission is in general aligned with the stellar and dust emission, indicating that molecular gas and dust are well mixed. Only for one object, Gal2, we find a different configuration: the dust emission, traced by the ALMA continuum, comes from two distinct regions, while detect CO only in the southernmost of the two (and we stress that the ALMA tuning could detect CO also in the north one, if CO were present and excited as in the south one).
    
    \item We find a positive correlation between $L'_{CO(5-4)}$ and $L_{FIR}$ for our five galaxies at $z\sim3$, confirming that the correlation that is observed at lower redshifts (Daddi et al. 2015; Liu et al. 2015) also holds for MS galaxies at $z\sim3$. The two galaxies that lie closer to the average MS at that redshift, that have also smaller SFR, are the ones with fainter  $L'_{CO(5-4)}$ luminosities. This findings support the claim that the CO(5-4) luminosity can be used by as an independent star-formation indicator, as suggested by Daddi et al. (2015).
    
    \item We find a correlation between three quantities: the CO SLED slope between CO(5-4) and CO(4-3), the $L'_{CO(5-4)}$ luminosity, and the distance from the MS $\delta_{MS}$ (see Figure~\ref{Fig:SLED_slope}: the two objects that lie closer to the MS have slopes similar to BzKs and SMGs, and have faint CO(5-4) luminosities. This indicates that the two former objects are likely normal MS galaxies that form stars in a secular mode, with large gas reservoirs and long gas depletion timescales (see Table \ref{tab:properties_alma2} and Figure~\ref{Fig:tdepl}). The other three, that lie closer to the starburst region, have definitely peculiar CO SLEDs: a possibility is that the molecular gas in these galaxies is in multiple phases, with the denser gas undergoing a very active star-forming episode, therefore emitting an excess of CO(5-4) photons with respect to the less dense gas.

    \item We find systematic differences in the molecular gas mass estimate when the CO SLED is used as opposed to the dust continuum in the Rayleigh-Jeans regime (Figure~\ref{Fig:COlum_FIR}): gas masses derived from the CO are larger (smaller) than dust based ones for galaxies with warmer (colder) dust temperature, and they agree for galaxies with dust temperatures T$\sim$54 K, as estimated by the MAGPHYS SED fitting.

    \item We also showed that constraining the slope of the CO SLED can help to at least reduce the uncertainties when extrapolating the SLED down to CO(1-0) to trace the molecular gas mass: by doing so, and then assuming a conversion factor $\alpha_{CO}=3.8~M_{\odot} (K~km~s^{-1}~pc^{-2})^{-1}$, typical for star-forming galaxies on the main sequence (Tacconi et al. 2018) we find that our five galaxies are very gas rich, with gas fractions between 60 and 80\%, values that are very close to the ones measured for similar (but lensed) galaxies at in the same redshift regime (Dessauges-Zavadsky et al. 2017). This puts on firmer grounds the findings in literature that the increase in gas fraction slightly flattens out at $z>3$ (Dessauges-Zavadsky et al. 2017; Schinnerer et al. 2016). On the other hand, we find   also that these galaxies have depletion times in the range $0.2<t_{depl}<1$ Gyr, again similar to the values found in literature for similar MS galaxies at z$\sim$3 (Dessauges-Zavadsky et al. 2017, but mainly for lensed galaxies). We also find that the two galaxies that lie close to the average MS have longer gas depletion timescales than the ones that lie at the boundary to starbursts: this suggests that the depletion times decrease moving up perpendicular to the main sequence, similar to what has been found by Schinnerer et al. (2016) using dust as a gas mass tracer.
    
    \item We obtain a first estimate of the dynamical masses, using common assumptions, that turn out to be comparable with the total {\it baryonic} mass (stellar+gas) in these galaxies, except for one galaxy, that is probably observed almost face-on. This is important, because demonstrates that the assumptions we made to constrain the molecular gas mass and the dynamical mass from the CO lines are reasonable. Moreover, it allows, {\it for the first time at these redshifts}, to put an observational constraint on the $\alpha_{CO}$ parameter, that turns out to be very close to the value of  $\alpha_{CO}=3.8~M_{\odot} (K~km~s^{-1}~pc^{-2})^{-1}$ prescribed by Tacconi et al. (2018) for normal star-forming galaxies.

\end{itemize}

\begin{acknowledgements}
 This work is based on ALMA data from the project ADS/JAO.ALMA \#2015.1.01590.S. ALMA is a partnership of ESO (representing its member states), NSF (USA), and NINS (Japan), together with NRC (Canada), NSC and ASIAA (Taiwan), and KASI (Republic of Korea), in cooperation with the Republic of Chile. The Joint ALMA Observatory is operated by ESO, AUI/NRAO, and NAOJ. We are grateful for the support from the italian regional ALMA ARC. PC and LM acknowledge support from the BIRD 2018 research grant from the Universit\`a degli Studi di Padova; PC acknowledges support from the CONICYT/FONDECYT program N$^\circ$\,1150216; EI acknowledges partial support from FONDECYT through grant N$^\circ$\,1171710; DL and ES acknowledge funding from the European Research Council (ERC) under the European Union's Horizon 2020 research and innovation programme (grant agreement No. 694343). We thank the anonymous referee for a helpful report that improved the clarity of the paper.
\end{acknowledgements}

%-------------------------------------------------------------------


\begin{thebibliography}{}
\bibitem[Bolatto et al. (2013)]{Bolatto13} Bolatto, A. D., Wolfire, M., Leroy, A. K., 2013, ARA\&A, 51, 207
\bibitem[Bothwell et al. (2013)]{Bothwell13} Bothwell M. S., Smail I., Chapman S. C., et al., 2013, MNRAS, 429, 3047
\bibitem[Briggs (1995)]{Briggs95} Briggs D. S., 1995, AAS Meeting Abstracts, p. 1444
\bibitem[Capak et al. (2007)]{capak07} Capak, P., Aussel, H., Ajiki, M., et al., 2007, ApJS, 172, 99 
\bibitem[Capak et al. (2015)]{capak15} Capak, P., Carilli, C., Jones, G., et al., 2015, Nature , 522, 455 
\bibitem[Carilli~\&~Walter~(2013)]{CW13} Carilli, C., \& Walter, F., 2013, ARA\&A, 51, 105
\bibitem[Combes et al. (2016)]{Combes16} Combes F., the PHIBSS collaboration 2016, Proc. IAUSymp. 315, From Interstellar Clouds to Star-Forming Galaxies: Universal Processes?, Cambridge Univ. Press, Cambridge, p. 240
\bibitem[Cucciati et al. (2012)]{cucciati} Cucciati, O., Tresse, L., Ilbert, O., et al., 2012, A\&A, 539, 31
\bibitem[da Cunha, charlot, \& Elbaz (2008)]{daCunha08} da Cunha E., Charlot S., Elbaz D., 2008, MNRAS, 388, 1595
\bibitem[daCunha et al. (2010)]{daCunha10} da Cunha E., Charmandaris, V., D\'iaz-Santos, T., et al., 2010, A\&A, 523, 78
\bibitem[Daddi et al. (2007)]{Daddi07} Daddi, E., Dickinson, M., Morrison, G., et al., 2007, ApJ, 670, 156
\bibitem[Daddi et al. (2010)]{Daddi10} Daddi, E., Bournaud, F., Walter, F., et al. 2010, ApJ, 713, 686
\bibitem[Daddi et al. (2015)]{Daddi15} Daddi, E., Dannerbauer, H., Liu, D., et al. 2015, A\&A, 577, 46
\bibitem[Dessauges-Zavadsky et al. (2015)]{dessauges15} Dessauges-Zavadsky, M., Zamojski, M., Schaerer, D., et al., 2015, A\&A, 557, 50
\bibitem[Dessauges-Zavadsky et al. (2017)]{dessauges17} Dessauges-Zavadsky, M., Zamojski, M., Rujopakarn, W., et al., 2017, A\&A, 605, 81
\bibitem[Elbaz et al. (2018)]{Elbaz18} Elbaz, D., Leiton, R., Nagar, N., et al., 2018, A\&A, 616, 110
\bibitem[Faisst et al. (2016)]{Faisst16} Faisst, A. L., Capak, P., Hsieh, B. C., 2016, ApJ, 821, 122
\bibitem[Geach et al. (2011)]{Geach11} Geach, J. E., Smail, I., Moran, S. M., et al., 2011, ApJ, 730, 19
\bibitem[Genzel et al. (2015)]{Genzel15} Genzel, R., Tacconi, L. J., Lutz, D., et al., 2015, ApJ, 800, 20
\bibitem[Genzel et al. (2017)]{Genzel17} Genzel, R., F\"orster-Schreiber, N. M., \"Ubler, H., et al., 2017, Nature, 543, 397
\bibitem[Greve et al. (2005)]{Greve05} Greve, T.R., Bertoldi, F., Smail, I., et al., 2005, MNRAS, 359, 1165
\bibitem[Groves et al. (2015)]{Groves15} Groves, B. A., Schinnerer, E., Leroy, A., et al., 2015, ApJ, 799, 96
\bibitem[Hildebrand (1983)]{Hildebrand83}Hildebrand, R. H., 1983, QJRAS, 24, 267
\bibitem[Jin et al. (2018)]{Jin18} Jin, S., Daddi, E., Liu, D., et al., 2018, ApJ, 864, 56
\bibitem[Kennicutt (1998)]{kennicutt98} Kennicutt, R.C. Jr., 1998, ARA\&A, 36, 189
\bibitem[Koekemoer et al. (2007)]{koekemoer07} Koekemoer, A. M., Aussel, H., Calzetti, D., et al., 2007, ApJS, 172, 196
\bibitem[Laigle et al. (2016)]{Laigle2016} Laigle, C., McCracken, H. J., Ilbert, O., et al., 2016, ApJS, 224, 24
\bibitem[Lang et al. (2017(]{Lang17} Lang, P., F\"orster Schreiber, N. M., Genzel, R., et al., 2017, ApJ, 840, 92
\bibitem[Le F\`evre et al. (2015)]{lefevre} Le F\`evre, O., Tasca, L. A. M., Cassata, P., et al., 2015, A\&A, 576, 79
\bibitem[Lilly et al. (2007)]{lilly07} Lilly, S. J., Le F\`evre, O., Renzini, A., et al., 2007, ApJS, 172, 70
\bibitem[Liu et al. (2015)]{Liu15} Liu, Daizhong; Gao, Yu; Isaak, K., et al., 2015, ApJ, 810, 14
\bibitem[Liu et al. (2019)]{Liu19} Liu, D., Schinnerer, E., Groves, B., et al., 2019, ApJ, [arXiv:1910.12883]
\bibitem[Madau \& Dickinson (2014)]{MD14} Madau \& Dickinson, 2014, ARA\&A, 52, 415 al., 2012, A\&A, 539, 31
\bibitem[Magdis et al. (2012a)]{Magdis12a}Magdis, G. E., Daddi, E., B\'ethermin, M., et al., 2012a, ApJ, 760, 6
\bibitem[Magdis et al. (2012b)]{Magdis12b}Magdis, Georgios E.; Daddi, E.; Sargent, M., et al., 2012b, ApJ, 758, 9
\bibitem[Magnelli et al. (2012)]{Magnelli12} Magnelli, B., Saintonge, A., Lutz, D., et al., 2012, A\&A, 548, 22
\bibitem[Mc Mullin et al. (2007)]{McMullin07} McMullin, J. P., Waters, B., Schiebel, D., et al., 2007, in Shaw R. A., Hill F., Bell D. J., eds, Astronomical Data Analysis Software and Systems XVI, Vol. 376. Astronomical Society of the Pacific, San Francisco, CA, p. 127
\bibitem[Narayanan~\&~Krumholz~(2014)]{Narayanan14} Narayanan, D., \& Krumholz, M. R., 2014, \mnras, 442, 1411
\bibitem[Noeske et al. (2007)]{Noeske07}Noeske, K. G., Weiner, B. J., Faber, S. M., et al. 2007, ApJ, 660, 43
\bibitem[Papadopulos et al. (2012)]{Papadopulos12} Papadopoulos, P.P., van der Werf, P., Xilouris, E., et al., 2012, ApJ, 751, 10
\bibitem[Rodighiero et al. (2011)]{Rodighiero11} Rodighiero, G., Daddi, E., Baronchelli, I., et al., 2011, ApJL, 739, 40
\bibitem[Rodighiero et al. (2014)]{Rodighiero14} Rodighiero, G., Renzini, A., Daddi, E., et al., 2014, \mnras, 443, 19
\bibitem[Saintonge et al. (2013)]{Saintonge13} Saintonge, A., Lutz, D., Genzel, R., et al., 2013, ApJ, 778, 2
\bibitem[Sanders et al. (2007)]{Sanders07} Sanders, D. B., Salvato, M., Aussel, H., et al., 2007, ApJS, 172, 86
\bibitem[Sargent et al. (2014)]{Sargent14} Sargent, M., Daddi, E., B\'ethermin, M., et al., 2014, ApJ, 793, 19
\bibitem[Schinnerer et al. (2016)]{Schinnerer16} Schinnerer, E., Groves, B., Sargent, M. T., et al., 2016, 833, 112
\bibitem[Schreiber et al. (2016)]{Schreiber16}Schreiber, C., Elbaz, D., Pannella, M., et al., 2016, A\&A, 589, 35
\bibitem[Sharon et al. (2016)]{Sharon16} Sharon, C. E., Riechers, D. A., Hodge, J., et al., 2016, ApJ, 827, 18
\bibitem[Scoville et al. (2007)]{Scoville07} Scoville, N., Aussel, H., Brusa, M., et al., 2007, ApJS, 172, 1
\bibitem[Scoville et al. (2014)]{Scoville14}Scoville, N., Aussel, H., Sheth, K., et al., 2014, ApJ, 783, 84
\bibitem[Scoville et al. (2016)]{Scoville16}Scoville, N., Sheth, K., Aussel, H., et al., 2016, ApJ, 820, 83
\bibitem[Tacconi et al. (2006)]{Tacconi06} Tacconi, L.J., Neri, R., Chapman, S.C., et al., 2006, ApJ, 640, 228
\bibitem[Tacconi et al. (2008)]{Tacconi08} Tacconi, L.J., Genzel, R., Smail, I., et al., 2008, ApJ, 680, 246
\bibitem[Tacconi et al. (2010)]{Tacconi10} Tacconi, L.J., Genzel, R., Neri, R., et al., 2010, Nature, 463, 781
\bibitem[Tacconi et al. (2013)]{Tacconi13} Tacconi, L.J., Neri, R., Genzel. R., et al., 2013., ApJ, 768, 74
\bibitem[Tasca et al. (2015)]{Tasca15} Tasca, L., Le F\`evre, O., Hathi, N. P., et al., 2015, A\&A, 581, 54
\bibitem[Tomczak et al. (2016)]{Tomczak16} Tomczak, A. R., Quadri, R. F., Tran, K.-V. H., et al., 2016, ApJ, 87, 118
\bibitem[Villanueva et al. (2017)]{Villanueva17} Villanueva, V., Ibar, E., Hughes, T. M., et al., 2017, MNRAS, 470, 3775
\bibitem[Whitaker et al. (2012)]{whitaker12} Whitaker, K. E., van Dokkum, P. G., Brammer, G., and Franx, M., 2012, ApJ, 754, 29

    
    

\end{thebibliography}
\end{document}